\documentclass[a4paper,onecolumn,11pt,accepted=2026-03-23]{quantumarticle}
\pdfoutput=1
\usepackage[utf8]{inputenc}
\usepackage[english]{babel}
\usepackage[T1]{fontenc}
\usepackage{amsmath}
\usepackage{hyperref}

\usepackage{tikz}
\usepackage{lipsum}

\usepackage{graphicx}
\usepackage{tabularx}
\usepackage{braket}
\usepackage{amssymb}
\usepackage{amsmath}
\usepackage{mathrsfs} 
\usepackage{bm}
\usepackage{enumerate}
\usepackage{eczoo}

\newcommand{\mN}{\mathcal{N}}
\newcommand{\mF}{\mathcal{F}}

\newcommand{\sL}{\mathscr{L}}
\newcommand{\sH}{\mathscr{H}}

\newcommand{\vvv}[1]{\ensuremath{\mathbf{#1}}}
\newcommand{\vh}{\vvv{h}}
\newcommand{\vL}{\vvv{L}}
\newcommand{\vK}{\vvv{K}}
\newcommand{\frakh}{\mathfrak{h}}

\newcommand{\lr}[1]{\left( #1\right)}
\newcommand{\mlr}[1]{\left[ #1\right]}

\newcommand{\alr}[1]{\left\langle #1\right\rangle}
\newcommand{\norm}[1]{\left\lVert#1\right\rVert}
\newcommand{\abs}[1]{\left\lvert#1\right\rvert}

\newcommand{\ii}{\mathrm{i}}
\newcommand{\ee}{\mathrm{e}}

\newcommand{\tr}[1]{\mathrm{tr}\lr{#1}}

\newcommand{\where}{\quad {\rm where}\quad}
\newcommand{\order}{\mathrm{O}}

\newcommand{\GHZ}{\mathrm{GHZ}}

\newcommand{\fY}{\mathbf{Y}}
\newcommand{\id}{\mathrm{id}}

\newcommand{\cL}{\mathcal{L}}

\newcommand{\DD}{{\mathcal{D}}}

\newcommand{\tZ}{\widetilde{Z}}
\newcommand{\G}{M}
\newcommand{\cF}{{\mathcal{F}}}
\newcommand{\OO}{{\mathcal{O}}}
\newcommand{\mytag}[1]{\tag*{\llap{$\lr{#1}$}}}

\usepackage{amsthm}
\newtheorem{thm}{Theorem}

\newtheorem{lem}[thm]{Lemma}

\newtheorem{prop}[thm]{Proposition}

\renewcommand{\eqref}[1]{\hyperref[#1]{Eq.~(\ref{#1})}}
\newcommand{\figref}[1]{\hyperref[#1]{Fig.~\ref{#1}}}
\newcommand{\thmref}[1]{\hyperref[#1]{Theorem~\ref{#1}}}
\newcommand{\tabref}[1]{\hyperref[#1]{Table~\ref{#1}}}

\newcommand{\revise}[1]{{#1}}

\begin{document} 

\title{Metrologically optimal quantum states under noise}
\author{Chao Yin}\email{chao.yin@colorado.edu}
\affiliation{Department of Physics and Center for Theory of Quantum Matter, University of Colorado, Boulder CO 80309, USA}

\author{Victor V. Albert}
\email{vva@umd.edu}
\affiliation{Joint Center for Quantum Information and Computer Science, NIST/University of Maryland, College Park, Maryland, USA}

\author{Sisi Zhou}\email{sisi.zhou26@gmail.com}
\affiliation{Perimeter Institute for Theoretical Physics, Waterloo, ON, N2L 2Y5, Canada}
\affiliation{Department of Physics and Astronomy, Department of Applied Mathematics, and Institute for Quantum Computing, University of Waterloo, Ontario N2L 2Y5, Canada}

\begin{abstract}
We propose a class of metrological resource states whose quantum Fisher information scales optimally in both system size and noise rate.
In these states, qubits are partitioned into sensing groups with relatively large correlations within a group but small correlations between groups.
The states are obtainable from local Hamiltonian evolution,
and we design a metrologically optimal and efficient measurement protocol utilizing time-reversed dynamics and single-qubit on-site measurements. 
Using quantum domino dynamics, we also present a protocol free of the time-reversal step that has an estimation error roughly twice the best possible value. Finally, we show that spin squeezed states
are also optimal for noisy metrology under general conditions.
\end{abstract}


\maketitle

\tableofcontents

\section{Introduction}
Quantum metrology~\cite{metro_rev11,sensing_rmp,pezze2018quantum,pirandola2018advances} is the study of utilizing entanglement and other quantum effects to 
measure unknown data, such as gravitational waves~\cite{caves1981quantum,yurke19862,ligo2011gravitational,ligo2013enhanced}, images~\cite{le2013optical,lemos2014quantum,tsang2016quantum,abobeih2019atomic}, magnetic fields~\cite{wineland1992spin,bollinger1996optimal,leibfried2004toward,taylor2008high,zhou2020quantum}, and time~\cite{rosenband2008frequency,appel2009mesoscopic,ludlow2015optical,kaubruegger2021quantum,marciniak2022optimal}. 
A typical metrological protocol involves initializing \(N\) probes in a particular (possibly entangled) state, letting the value of an unknown quantity \(\theta\) be imparted on the state, and estimating \(\theta\) via a measurement.
The Heisenberg limit (HL)~\cite{giovannetti2006quantum}
restricts the best-case estimation precision to scale inversely with the number \(N\) of entangled probes, 
$\delta \theta \sim 1/N$.

The prototypical example of a ``metrologically optimal'' state is the Greenberger-Horne-Zeilinger (GHZ) state~\cite{giovannetti2006quantum}, which is used to attain the HL in estimating a $Z$-axis magnetic field in a noiseless \(N\)-spin system.
Unfortunately, the useful but fragile quantum correlations of this state quickly break down in the presence of noise.
Instead of attaining the HL, the optimal scaling 
in a generically noisy setting is the standard quantum limit (SQL), $\delta \theta \sim 1/\sqrt{N}$~\cite{giovannetti2006quantum,metro_noise12,metro_noise11,demkowicz2014using,sekatski2017quantum,metro_QEC17,zhou2018achieving}. 

For quantum metrology to continue to bear fruit, it is important to quantify the performance of noise-robust metrological protocols not only in terms of the number of probes \(N\)~\cite{giovannetti2006quantum,metro_noise12,metro_noise11,demkowicz2014using,sekatski2017quantum,metro_QEC17,zhou2018achieving}, \textit{but also} in terms of the noise rate \(p\) of each probe. 
A more precise calculation for several models with non-negligible noise yields a precision of $\delta \theta \gtrsim \sqrt{p/N}$ ~\cite{huelga1997improvement,metro_noise12,metro_noise11,demkowicz2014using,channel_estimate21}, meaning that reducing the noise rate \(p\) can lead to a $\sqrt{p}$ improvement on top of the naive scaling with \(1/\sqrt{N}\). While the improvement can be significant for a small $p$, it was rarely explored previously what types of quantum states and measurement protocols can achieve it. Numerical evidence~\cite{MPS_metro13,MPS_metro20} suggests that matrix product states with low bond dimension may be optimal for strong noise. However, these results are based on specific examples, without a general theoretical framework, and the underlying mechanisms behind this behavior remain unclear.

\begin{figure}
    \centering
    \includegraphics[width=.8\textwidth]{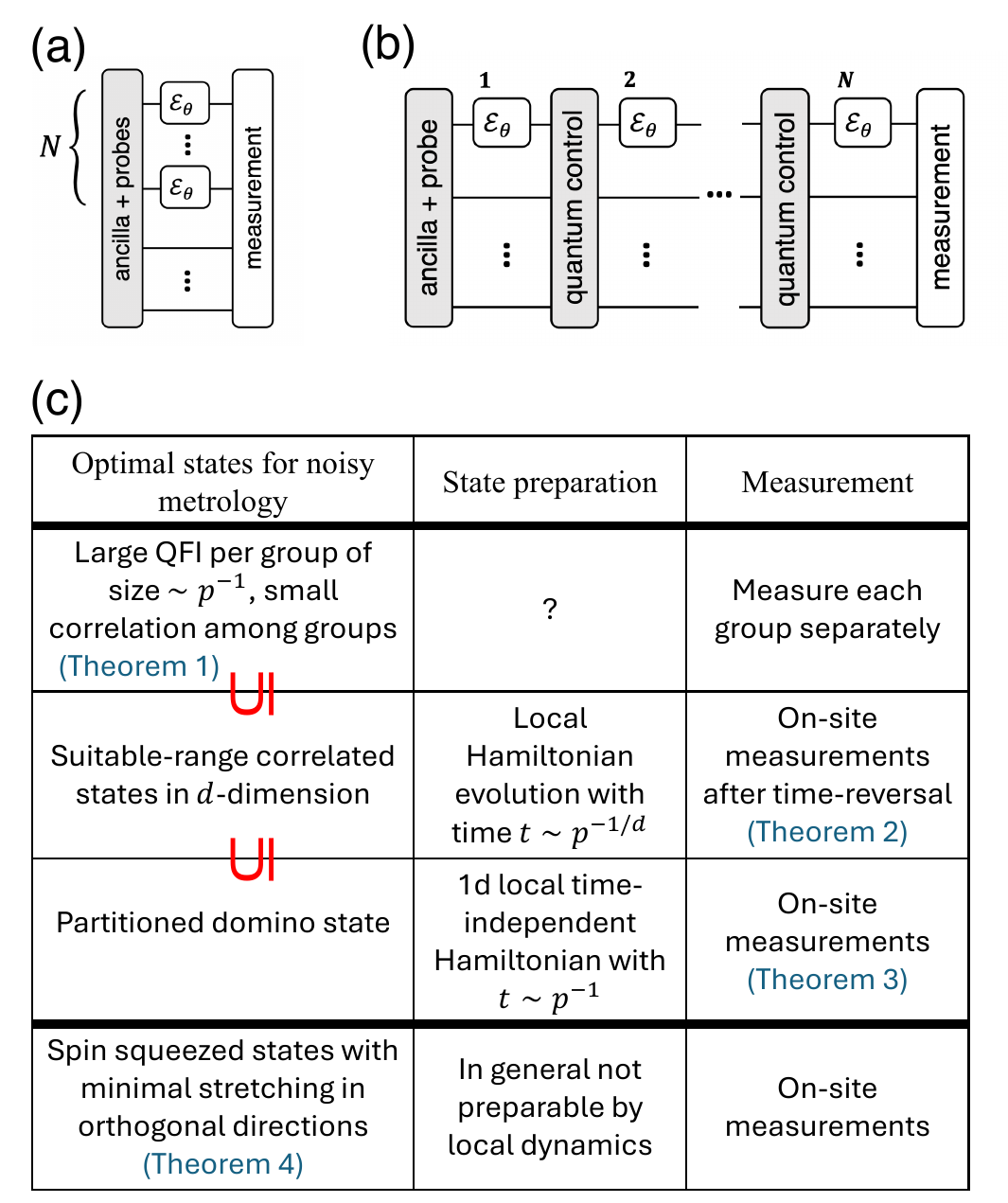}
    \caption{
    \revise{Quantum metrology can be done in either parallel (a) or sequential (b) strategies. Although examples are known for both strategies that have the optimal performance under noisy metrology in terms of scaling \cite{demkowicz2014using,channel_estimate21}, it has remained unclear whether the resource states used in parallel strategies extend beyond isolated examples to form general and systematic constructions. In this work, we find (c) four \emph{families}} of metrological optimal states and their corresponding preparation and measurement protocols. The first three satisfy the small correlation condition~\eqref{eq:sumJCIJ<'} \revise{that generalizes the quantum parity state example}, and
    each row is a special case of the previous row as indicated by {\color{red}$\subseteq$}.
    Each nested subset admits a more efficient scheme for state preparation and measurement than the previous one.
    Spin-squeezed states, described in the fourth row, violate~\eqref{eq:sumJCIJ<'} and do not belong to the previous families.}
    \label{fig:summary}
\end{figure}

In this work, we pioneer the study of quantum estimation limits as functions of both $N$ and $p$. We successfully identify criteria on and examples of protocols whose variance scales optimally~\footnote{We say a family of quantum states $\rho$ as functions of $N$ and $p$ is \emph{optimal}, if and only if for any $N$ and $p$, there exists a universal constant $c$ such that the QFI $\cF(\rho_\theta) \geq c \min\{N/p,N^2\}$.} with $N$ and $p$ and also construct efficient state preparation and measurement protocols. See \figref{fig:summary} for a summary. The extensive entanglement of the GHZ state is optimal in the relatively noiseless setting of \(pN\sim\) constant~\cite{hayashi2022global}, while unentangled states do not yield any \(p\)-dependent precision improvement.
Interpolating between the two edge cases, we show that a ``Goldilocks'' amount of quantum correlation---quantified by a correlation length \(M \sim 1/p\)---is optimal. We also establish the optimality criteria on a different class of states that may be long-range correlated: spin squeezed states. Throughout, we rigorously prove our results in Appendix, and provide intuitions and proof sketches in the main text. \revise{Although we are primarily interested in saturating the asymptotic scaling $\delta\theta\sim \sqrt{p/N}$ rather than the numerical prefactor, we will show examples with good quantitative performance.}


\section{Setting and a Bound on quantum Fisher information}
In this work, we will state our optimality conditions for sensing a \(Z\)-axis magnetic field under arbitrary noise, with a straightforward extension to any single-qudit Hamiltonian rotations presented in Appendix. 

We use dephasing noise as an illuminating warm-up example.
Let $N$ qubits, labeled by $i\in\Lambda$ with Pauli matrices $X_i,Y_i,Z_i$, be prepared in a state $\rho$, with each qubit $i$ undergoing the channel \begin{equation}\label{eq:Domega}
    \DD_{\theta, p}(\rho_i) = (1-p)\ee^{-\ii \theta Z_i} \rho_i \ee^{\ii \theta Z_i} + p Z_i\ee^{-\ii \theta Z_i} \rho_i \ee^{\ii \theta Z_i}Z_i
\end{equation}
for $0 < p < 1/2$. 
For simplicity, we assume $p > 1/N$ (otherwise the noise is asymptotically negligible and the HL is achievable).
One then measures the final state $\rho_\theta = \DD_{\theta, p}^{\otimes N}(\rho)$
to estimate $\theta$. 

The estimation precision can be bounded by the state's quantum Fisher information (QFI), $\delta \theta \geq 1/\sqrt{N_{\rm e}\cF(\rho_\theta)}$, via the quantum Cram\'{e}r-Rao bound~\cite{QFI76,holevo2011probabilistic,QFI94,barndorff2000fisher,gill2000state,estimation_tech09}, saturable as the number of repeated experiments $N_{\rm e} \rightarrow \infty$.
The QFI for the above example is bounded by
\setcounter{thm}{-1}
\revise{
\begin{thm}[Bound on QFI ~\cite{huelga1997improvement,metro_noise11,metro_noise12,channel_estimate21} ]
    \begin{equation}\label{eq:optimal}
    \cF(\rho_\theta)\le 4N/[1-(1-2p)^2] \sim N/p.
\end{equation}
\end{thm}
}

\revise{This yields} the aforementioned scaling of \(\delta\theta\sim\sqrt{p/N}\)~\footnote{We use $f\sim g,f\gtrsim g$ and $f\lesssim g$ to denote $c_1 g \leq f \leq c_2 g, f \geq c_3 g$ and $f \leq c_4 g$, respectively, where $c_{1,2,3,4}$ are positive constants.}, which also applies to general noise channels. 
This QFI bound cannot be surpassed with arbitrary state preparation and measurement protocols, including quantum error correction~\cite{sekatski2017quantum,zhou2018achieving,metro_QEC17}. 

\revise{Note that \eqref{eq:optimal} cannot be attained by separable states, for which $\cF(\rho_\theta) \leq 4N$ 
scales} constantly with $p$. The optimal product state is $\rho = \left(\frac{\ket{0}+\ket{1}}{\sqrt{2}}\frac{\bra{0}+\bra{1}}{\sqrt{2}}\right)^{\otimes N}$, which achieves $\cF(\rho_\theta) = 4(1-2p)^2 N$. This clearly has a gap between the $\sim N/p$ upper bound for small $p$. 

\section{An optimal example: quantum parity state}

\revise{The QFI upper bound \eqref{eq:optimal} is the optimal scaling with respect to both $N$ and $p$, because it can be saturated (up to a constant factor) by the following simple example: }
Consider the quantum parity state, \revise{\begin{equation}
    \rho = \bigotimes_{I} \ket{\GHZ}_I \bra{\GHZ}_I
\end{equation}
}with $\ket{\GHZ}_I = \frac{1}{\sqrt{2}}(\ket{0}^{\otimes \G} + \ket{1}^{\otimes \G})$ (an instance of the \eczoo[quantum parity code]{quantum_parity} \cite{Shor_code95,knill_paritycode00,paritycode05}). \revise{Here we have partitioned the \(N\)-qubit system into \(N/M\) groups labeled by $I\subset \Lambda$, each containing \(M\) qubits (with $N/\G$ assumed to be an integer for simplicity). We do not want to have the global GHZ state $M=N$ because its quantum entanglement is destroyed for any $p\gg 1/N$. }
To find a suitable choice of $M$, we first note that 
in the noiseless limit $p=0$, each group $I$ gains a phase $\G\theta$ independently that corresponds to a QFI $\sim \G^2$, so the QFI of the entire state, equal to the sum of individual QFIs, scales as $(N/\G)\G^2=N\G$. In addition, this scaling is robust against noise if $p \G$ is upper bounded by a constant, because then dephasing noise will only cause the individual QFI to reduce by a constant factor~\cite{huelga1997improvement}. Therefore, \eqref{eq:optimal} is saturated by tuning $\G \sim 1/p$ (see also \cite{zang2025enhancing}). \revise{To see these claims concretely, \begin{align}
    \mathcal{F}(\rho_\theta) &= \frac{N}{M}\mathcal{F}(\rho_{I,\theta}) \mytag{\rho_\theta=\bigoplus_I \rho_{I,\theta};\text{Additivity of QFI}} \nonumber\\
    &= \frac{N}{M} 4 \sum_{a<b}\frac{(\lambda_a-\lambda_b)^2}{\lambda_a+\lambda_b} \abs{\alr{a|Z_I|b}}^2 \hspace{10em}\mytag{Z_I:=\sum_{i\in I} Z_i} \\
    &=4NM(1-2p)^{2M}. \label{eq:QFI_GHZ}
\end{align}
Here the second line uses the definition of QFI, where $\{(\lambda_a,\ket{a})\}$ are eigenvalues and eigenstates of the density matrix in group $I$ \begin{equation}
    \rho_{I,\theta} = \frac{1}{2}\left(\begin{array}{cc}
        1 & (1-2p)^M \\
        (1-2p)^M & 1
    \end{array}\right),
\end{equation}
(setting $\theta=0$) supported in the subspace containing $\ket{0\cdots 0},\ket{1\cdots 1}$. The eigenstates are therefore $(\ket{0\cdots 0}\pm\ket{1\cdots 1})/\sqrt{2}$ with $\lambda_a+\lambda_b=1$ and $|\lambda_a-\lambda_b|=(1-2p)^M$, which leads to the last line of \eqref{eq:QFI_GHZ} using $\alr{a|Z_I|b}=M$. At $p\ll1$ ($p$ does not need to scale with $N$), \begin{equation}\label{eq:GHZ_e}
    \mathcal{F}\lr{\rho_{\theta}} =4NM(1-2p)^{2M}\approx 4NM \ee^{-4Mp}\approx \frac{N}{\ee p},
\end{equation}
saturates the optimal scaling \eqref{eq:optimal} up to a factor $1/\ee$, where we have
maximized the expression at $M\approx 1/(4p)$.}

Replacing the GHZ state in each group by another state that achieves the HL within the group, such as a spin-squeezed state~\cite{squeez_93,squeez_rev11,ulam2001spin,brask2015improved,sorensen2001entanglement}, does not change that arguments above and yields another optimal state. 

The state above has a tensor-product structure w.r.t.\@ the partitioning into groups. This may be too strict of a condition for practical purposes, e.g., when nearest neighbor interactions in the quantum sensor cannot be turned off. Therefore, it is crucial to understand the \emph{robustness} of this partitioning strategy in the presence of correlations among the groups. This is a nontrivial task since the QFI of noisy mixed states is difficult to analyze and has no obvious relation with usual correlation measures \cite{metro_rev14}. We solve this problem below by showing that any state with sufficiently large intra-group and sufficiently small inter-group correlations achieves the optimal QFI scaling $\sim N/p$. Moreover, our results throughout hold for general noise channels: The final state is $\rho_\theta=\mathcal{N}_{\theta, p}^{\otimes N}(\rho)$, where $\mathcal{N}_{\theta, p}$ is a general single-qubit noisy channel depending on $\theta$, and the noise parameter \(p\) quantifies 
the diamond distance between noiseless ($\mathcal{N}_{\theta, 0}$) and noisy ($\mathcal{N}_{\theta, p}$) evolutions (see Appendix \ref{sec:generalnoise} for precise definitions).

\revise{As a remark of the scope of this work, we focus on parallel strategies for quantum metrology \cite{demkowicz2014using}, where an entangled state is sent into the sensing channel once. The signal can also be sensed multiple times where coherence of the quantum register could provide quantum advantage (see \figref{fig:summary}). The quantum parity state example above can be mapped to an optimal sequential protocol using one qubit, which saturates the scaling \eqref{eq:optimal} where now $N$ means the number of sensing channels. The sequential protocol is simply letting the qubit coherently sense as many $\sim 1/p$ channels as possible set by the decoherence rate $p$. However, there are situations where one would prefer a parallel strategy rather than a sequential one, e.g. when saving the total sensing time, which motivates this work on \emph{robust} families of parallel strategies. }

\section{Sufficient condition for optimality}
Before stating our general sufficient condition (see Appendix for the proof), we first introduce some notation.
For any state $\sigma$ and operators $A,B$, define the expectation $\alr{A}_\sigma:=\tr{\sigma A}$, the correlation function 
\begin{equation}
    C_\sigma(A,B):=\frac{1}{2}\alr{A B+BA}_\sigma - \alr{A}_\sigma\alr{B}_\sigma,
\end{equation}
and the variance $\alr{\Delta A^2}_\sigma=C_\sigma(A,A)$. 
Let $\rho_I$ be the reduced density matrix (RDM) of group $I$ before sensing, and let $\rho_{I,\theta}$ ($\rho^{(0)}_{I,\theta}$) be the noisy (noiseless) RDM after sensing.
We use $A_I$ to indicate that operator $A$ acts inside group $I$, and denote the operator norm by $\norm{\cdot}$. The correlation between two groups is $C_\sigma(I;J):=\max_{\norm{A_I}=\norm{A_J}=1} \abs{C_\sigma(A_I, A_J)}$.

\begin{thm}\label{thm1}
For any given $p$, choose any $\G<N$ with \begin{equation}\label{eq:p<G}
    \G\le c_{\rm no}/p,
\end{equation}
where $c_{\rm no}$ is a positive constant, and partition the qubits into $N/M$ groups as above. Given any state $\rho$, suppose for any group $I$, its QFI in the noiseless limit is large, \begin{equation}\label{eq:FI0>G2}
        \cF\lr{\rho^{(0)}_{I,\theta}} \sim \G^2,
    \end{equation}
    and its correlations with other groups are small
    \begin{equation}
    \textstyle\sum_{J: J\notin \mathcal{S}_I} C_\rho(I;J) \lesssim \G^{-2}, \label{eq:sumJCIJ<'}
\end{equation}
where $\mathcal{S}_I$ is some set of groups containing a finite (independent of $N,p,M$) number of elements (including $I$) that may have large correlation with $I$.
Then
there exists an observable $\OO=\sum_I \OO_I$ such that \begin{equation}\label{eq:F>O>NG}
    \cF(\rho_\theta)\ge \frac{1}{\alr{\Delta \OO^2}_{\rho_\theta}}\lr{\partial_\theta\alr{\OO}_{\rho_\theta}}^2 \sim  N\G.
\end{equation}
\end{thm}

Note that $\alr{\Delta \OO^2}_{\rho_\theta}$ and $\alr{\OO}_{\rho_\theta}$ represent its variance and expectation value w.r.t.\@ $\rho_\theta$, respectively. See, e.g., Ref.~\cite{pezze2009entanglement} for the lower bound on QFI in \eqref{eq:F>O>NG}. 
$\OO$ here give rise to an optimal observable to measure, which is a sum of individual observables $\OO_I$ on each group. It implies individual measurements on each group suffice to achieve the optimal QFI, which may still be challenging to implement in cases when the group size $M$ is large.

\thmref{thm1} implies the optimal QFI scaling (\eqref{eq:optimal}) is achieved for a family of states when $\G \sim 1/p$ and two conditions in \eqref{eq:FI0>G2} and \eqref{eq:sumJCIJ<'} are satisfied. 
They generalize the quantum parity state example: 
\begin{enumerate}[(i)]
    \item Large intra-group correlation: \eqref{eq:FI0>G2} requires that each group has the asymptotically optimal QFI $\sim M^2$~\footnote{\label{foot:cor}In Appendix, we show that when $\rho$ is pure, the QFI of each RDM can be interpreted as certain correlation functions of $\rho$.}. (For the quantum parity state, the RDM of each group is a $M$-qubit GHZ state, satisfying this condition.) 
    \item Small inter-group correlation: \eqref{eq:sumJCIJ<'} requires each group to have a large correlation with \emph{at most} a finite number of other groups. (For the quantum parity state, each group has zero correlations with others.)
\end{enumerate}
To prove \thmref{thm1}, we first use the large intra-group correlation condition to prove that the QFI is $\sim NM$ when $p = 0$, similar to the quantum parity state. Next, we use the small inter-group correlation condition to prove, when $M$ is sufficiently small, i.e., $M \sim 1/p$, that the QFI is only negligibly perturbed by noise. We also adopt similar proof ideas for all other theorems below, where we approach the problem by first calculating the noiseless case and then show negligible perturbation under noise. As a result, our theorems apply for general Hamiltonian and noise models.
As we will see below, our general criteria for optimal noisy metrology can be satisfied by a large family of many-body states, where efficient preparation and measurement protocols exist. 

\section{Locally generated optimal states}
We show \thmref{thm1} can be naturally satisfied by states generated by local Hamiltonian evolution. Here, local Hamiltonians are sums of geometrically local terms supported on a constant-size sphere on a lattice.

Suppose the qubits lie on vertices of a $d$-dimensional square lattice, start from $\ket{\bm{0}}:=\ket{0}^{\otimes N}$. We consider states $\ket{\psi} = U \ket{\bm{0}}$, where unitaries $U$ are generated by a (possibly time-dependent) local Hamiltonian for time $T$. Such evolution is \textit{quasi-causal} because it satisfies the following Lieb--Robinson bound~\cite{Lieb1972,review_us23}: if an operator $A$ acts within set $S\subseteq \Lambda$, then, in the Heisenberg picture,  \begin{equation}\label{eq:AT=}
    A(T):= U^\dagger A U \approx \widetilde{A}_{\widetilde{S}},
\end{equation}
i.e., \(A(T)\) can be approximated (up to small error) 
by an operator $\widetilde{A}_{\widetilde{S}}$. It is strictly supported in the ``light cone'' region \begin{equation}\label{eq:tildeS}
    \widetilde{S}:=\{i\in \Lambda: \mathsf{d}(i,S)\le T\},
\end{equation} 
This region includes all vertices within a distance $T$ from $S$. 
Note that here $T$ is rescaled such that the Lieb--Robinson velocity \cite{review_us23} is set to $1$, $\mathsf{d}(\cdot, \cdot)$ is the distance function with $\mathsf{d}(i,S)=\min_{j\in S}\mathsf{d}(i,j)$, and we will assume \eqref{eq:AT=} is exact from now on, with the rigorous treatment given in Appendix.

The state $\rho=\ket{\psi}\bra{\psi} =U \ket{\bm{0}}\bra{\bm{0}}U^\dagger$ has correlation length at most $2T$~\cite{LRB_cor06} because faraway light cones do not intersect. 
The condition \eqref{eq:sumJCIJ<'} is then satisfied if each group $I$ is chosen to be an $L\times \cdots \times L$ hypercube with $L=\G^{1/d}\gtrsim T$. Each group can have a large correlation with only its neighboring groups.

To satisfy the condition \eqref{eq:FI0>G2}, we first choose \(T \sim L\) so that qubits within each group are sufficiently correlated so as to satisfy \eqref{eq:FI0>G2}~\cite{LRB_metro23}. 
Then \eqref{eq:FI0>G2} is equivalent to the requirement of a large intra-group $ZZ$ correlation~\footnotemark[3], i.e., 
\begin{equation}\label{eq:C>LT}
    C_\rho(Z_I, Z_I) \ge c_Z L^d T^d,
\end{equation}
where $Z_I=\sum_{i\in I}Z_i$ and $c_Z \sim 1$.

We will give an example $\ket{\psi}=U\ket{0}$ that satisfies \eqref{eq:C>LT} later. For now, we consider \emph{any} optimal state that is efficiently preparable by local Hamiltonian evolution,
and for which we propose an efficient measurement protocol below. 

\section{On-site measurement scheme with time-reversal} 
For locally generated states, we propose an on-site measurement scheme by assuming the ability to time-reverse the evolution $U$ (e.g., by changing the sign of the generating Hamiltonian). 

After sensing, one applies a unitary $U_{\rm rev}:=U^\dagger \ee^{\ii \theta_{\rm pr} Z}$ to the state $\rho_{\theta}$, and then measures in the computational basis before post-processing the data. Here, $\theta_{\rm pr}$ represents prior knowledge of $\theta$, such that $\tilde{\theta}:=\theta-\theta_{\rm pr}$ satisfies $|\tilde\theta| \sim L^{-d}$~\footnote{The requirement $|\tilde{\theta}| \lesssim L^{-d}$ is a common assumption in metrology because one can always gain the prior knowledge on the ``digits'' of $\theta$ one by one with negligible overhead until the requirement is satisfied~\cite{bitwise_sense09,bitwise_sense15,bitwise_sense20}. We also assume $|\tilde{\theta}| \gtrsim L^{-d}$, achievable by shifting $\theta_{\rm pr}$ if necessary, so that the slope \eqref{eq:-dtheta=L2} is large.}.
The idea comes from the Loschmidt echo protocol working in the noiseless case~\cite{echo_metro16}, where one processes the measurement data by simply projecting the state to $\ket{\bm{0}}$. The probability is 
\begin{equation}\label{eq:Pecho}
    \mathsf{P}_{\rm echo}:=\abs{\bra{\bm{0}}U^\dagger \ee^{-\ii \tilde{\theta} Z} U\ket{\bm{0}} }^2 \!= 1 - \frac{\tilde{\theta}^2}{4}\! \mathcal{F}(\rho_\theta) + \order(\tilde{\theta}^4).
\end{equation} 
The projection is optimal because the corresponding classical Fisher information $\frac{1}{\Delta \mathsf{P}_{\rm echo}^2}\lr{\partial_\theta \mathsf{P}_{\rm echo}}^2 \approx \mathcal{F}(\rho_\theta)$ at small $\tilde{\theta}$, where $\Delta \mathsf{P}_{\rm echo}^2=\mathsf{P}_{\rm echo}(1-\mathsf{P}_{\rm echo})$ is the variance of a Bernoulli random variable with probability $\mathsf{P}_{\rm echo}$. 

However, the Loschmidt echo protocol above through measuring the observable $\ket{\bm{0}}\bra{\bm{0}}$ no longer works under noise. To guarantee robustness against noise, we need to post-process the data from computational-basis measurement in a different way. Let 
$P_{\widetilde{I}} := \bigotimes_{i\in \widetilde{I}}\ket{0}_i\bra{0}$ be projections onto small regions $\widetilde{I}$ (defined via \eqref{eq:tildeS}), we measure instead the overall observable $\alr{\OO_{\rm rev}}_{\rho_\theta}$ with 
\begin{equation}
   \textstyle\OO_{\rm rev} := \sum_I \widetilde{P}_{I} := \sum_I U_{\rm rev}^\dagger P_{\widetilde{I}} U_{\rm rev}.  
\end{equation}
Note that we often omit the identity operator on the rest of the system for simplicity, e.g., here $P_{\widetilde{I}}$ means $P_{\widetilde{I}} \otimes \id$. This time-reversal on-site measurement scheme is asymptotically optimal. 

\begin{thm}\label{thm:Orev>}
    For any $p$, choose $L=M^{1/d}$ such that \eqref{eq:p<G} holds for some constant $c_{\rm no}$ (which is not necessarily the same constant in Theorem \ref{thm1}). For any locally generated state $\rho$ satisfying \eqref{eq:C>LT}, there exist constants $C>c>1$ determined by $c_Z$ and $d$, such that
    for any $C\ge L/T\ge c$, \begin{equation}\label{eq:O>NL}
    \frac{1}{\alr{\Delta \OO^2_{\rm rev}}_{\rho_\theta}}\lr{\partial_\theta\alr{\OO_{\rm rev}}_{\rho_\theta}}^2 \sim  N L^d.
\end{equation}
\end{thm}
\thmref{thm:Orev>} implies locally generated optimal states, combined with the time-reversal measurement protocol, achieve the optimal QFI when choosing $M \sim 1/p$. 
An additional advantage of the time-reversal measurement protocol is the observable $\OO_{\rm rev}$ is independent of $p$, in contrast to $\OO$ in \thmref{thm1} that may be a function of $p$. 

Below we sketch the proof idea of \thmref{thm:Orev>}, leaving the full proof to Appendix. 
Note that all approximations ``$\approx$'' below stem from the light-cone approximation in \eqref{eq:AT=}.

\begin{figure}[htbp]
\centering
\includegraphics[width=.3\textwidth]{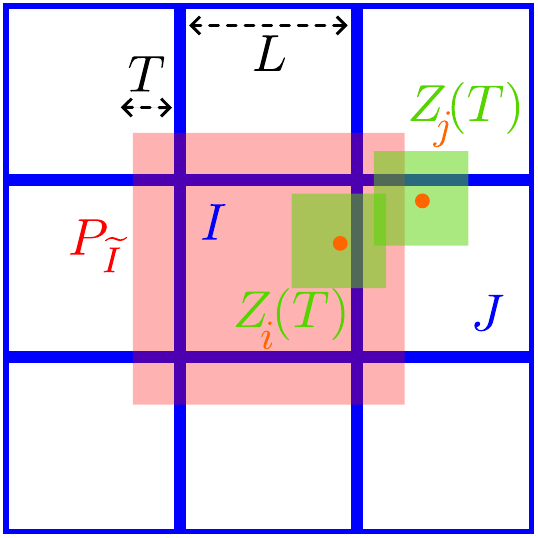}
\caption{\label{fig:ZI}
A coarse-grained sketch for the time-reversal on-site measurement protocol in 2d, where the underlying lattice is not shown. There is an observable $P_{\widetilde{I}}$ for each group $I$ after the time-reversal $U_{\rm rev}$, whose expectation is only affected by dynamics in the neighboring groups since $L> T$. As a crucial step to establish large sensitivity \eqref{eq:-dtheta=L2}, we show terms like $\bra{\bm{0}}[Z_I(T), [Z_J(T), P_{\widetilde{I}}]]\ket{\bm{0}}$ contribute subdominantly in \eqref{eq:PI=expand}, using the fact that only sites $i\in I, j\in J$ (orange dots) whose operators $Z_i(T)$ and $Z_j(T)$ (green regions) have overlapping support contribute to these terms. 
}
\end{figure}

Focusing on \(d=2\) (generalization is straightforward) as shown in \figref{fig:ZI} and recalling that \(|\tilde\theta|\sim L^{-2}\) and \(T\sim L\), we first expand in the noiseless case,
\begin{align}\label{eq:PI=expand}
    \langle\widetilde{P}_I\rangle_{\rho_\theta^{(0)}} &= \bra{\bm{0}} \ee^{\ii \tilde{\theta} Z(T)} P_{\widetilde{I}} \ee^{-\ii \tilde{\theta} Z(T)} \ket{\bm{0}} \nonumber\\
    &= 1-\frac{\tilde{\theta}^2}{2} \bra{\bm{0}}[Z(T), [Z(T), P_{\widetilde{I}}] ]\ket{\bm{0}} + \cdots.
\end{align}
The first-order term vanishes similarly as \eqref{eq:Pecho}, and we show that the second-order term is $\sim \tilde{\theta}^2 L^4$.
(Note that we have required that $|\tilde{\theta}|$ is sufficiently large to guarantee that the second-order derivative is dominant.)
We expand $Z=\sum_J Z_J$ and observe that only $J=I$ or neighbors of $I$ contribute, because otherwise $[Z_J(T), P_{\widetilde{I}}]=0$ since they do not overlap in space from \eqref{eq:AT=}. Furthermore, when $L/T$ is a sufficiently large constant, 
the coefficient is dominated by the term
\begin{equation}\label{eq:ZIZI>}
    \bra{\bm{0}}[Z_I(T), [Z_I(T), P_{\widetilde{I}}] ]\ket{\bm{0}} \approx 2\, C_\rho(Z_I, Z_I)\ge 2c_Z L^2T^2,
\end{equation}
which can be verified from direct computation and \eqref{eq:C>LT}. The reason is that the sum of all other terms is $\lesssim LT^3$, which becomes subdominant when $L/T$ is sufficiently large. To show this, we note that for $i\in I,j\in J\neq I$, \begin{equation}\label{eq:ZiZj}
    \bra{\bm{0}}[Z_i(T), [Z_j(T), P_{\widetilde{I}}] ]\ket{\bm{0}}\approx 0,
\end{equation}
unless the two sites are close ($\mathrm{d}(i,j)\le 2T$), as shown in \figref{fig:ZI}. 
Then the sum of \eqref{eq:ZiZj} over $i,j$ is bounded by $\sum_{j \text{ close to } I}\sum_{i \text{ close to } j} \lesssim LT \cdot T^2$.

Above we showed, at sufficiently large $L/T$, the second-order coefficient in \eqref{eq:PI=expand} is $\sim L^4$. This guarantees $-\partial_\theta\langle\widetilde{P}_I\rangle_{\rho_\theta^{(0)}} \sim \tilde{\theta} \times \text{[the second-order coefficient]} \sim L^2$, and thus \begin{equation}\label{eq:-dtheta=L2}
    \partial_\theta\alr{\OO_{\rm rev}}_{\rho_\theta^{(0)}}\sim N. 
\end{equation}
(Note that the higher orders of $\tilde\theta$ in \eqref{eq:PI=expand} are safely ignored by similar locality analysis.) Since faraway groups are hardly correlated, 
\begin{equation}\label{eq:DeltaO<}
    \alr{\Delta \OO^2_{\rm rev}}_{\rho_\theta^{(0)}} \sim \sum_I \big\langle\Delta \widetilde{P}_I^2\big\rangle_{\rho_\theta^{(0)}} \lesssim N/L^2,
\end{equation}
leading to \eqref{eq:O>NL} for the noiseless case.

Finally, we consider the noisy case $p>0$. \eqref{eq:DeltaO<} still holds (where $\rho_\theta^{(0)}$ is replaced with $\rho_\theta$) because adding noise does not create long-range correlation~\cite{LRB_open10}. 
It then remains to show \eqref{eq:-dtheta=L2} is also robust. For illustrative purposes, we focus on dephasing noise here, i.e., \(\mathcal{N}_{\theta,p}=\DD_{\theta,p}\), yielding
\begin{align}\label{eq:P-P0<}
    &\abs{ \partial_\theta \big( \big\langle{\widetilde{P}_I}\big\rangle_{\rho_\theta}- \big\langle{\widetilde{P}_I}\big\rangle_{\rho_\theta^{(0)}} \big)} = \abs{ \big\langle{ \big[ Z,\DD_{0, p}^{\otimes N}(\widetilde{P}_I)-\widetilde{P}_I } \big]  \big\rangle_{\rho_\theta^{(0)}} } \nonumber\\
    & \lesssim L^2   \norm{\DD_{0, p}^{\otimes c_0L^2}(\widetilde{P}_I)-\widetilde{P}_I} \lesssim L^4 p,
\end{align}
where we recall that the dual map of the dephasing noise $\DD_{0, p}$ 
is itself, and use the locality of $\widetilde{P}_I$ to restrict the channel to a region of size $c_0 L^2$ with $c_0 \sim 1$. 
Taking $p L^2 \leq c_{\rm no}$ and $c_{\rm no}$ 
sufficiently small guarantees the contribution from noise to \eqref{eq:-dtheta=L2} is subdominant relative to the noiseless result. This completes the proof sketch of Theorem \ref{thm:Orev>}.

\section{Quantum domino dynamics with on-site measurement}
We have shown that any state generated by local dynamics with correlations of the form \eqref{eq:C>LT} is metrologically optimal when $L^{d} \sim 1/p$. Here we present an explicit example of this framework using quantum domino dynamics~\cite{domino05,metro_domino21}, where an on-site measurement protocol without time-reversal is sufficient to achieve the optimal QFI. 

Consider the $3$-local spin-chain Hamiltonian that flips a spin only if its two neighbors are anti-aligned, \begin{equation}\label{eq:H=XZZ}
    H_{\rm do} = \frac{1}{2}\sum_{i=2}^{N-1} X_i(1-Z_{i-1}Z_{i+1}),
\end{equation}
which can be engineered as the effective Hamiltonian for the transerve-field Ising model at weak $X$ field \cite{domino05}.
The vacuum $\ket{\bm{0}}$ is an eigenstate of $H_{\rm do}$, and, more generally, the number of domain walls (DWs)---transitions from $\ket{0}$ to $\ket{1}$ or vice versa---is conserved. If the system starts from the initial state $\ket{1_1}:=\ket{10\cdots 0}$, it only traverses the state space with one DW $\ee^{-\ii T H_{\rm do}} \ket{1_1} = \sum_{i=1}^N a_i \ket{1_i}$, where $\ket{1_i}$ is the state with $1$ on the first $i$ qubits and $0$ on the rest and the coefficients $a_i$ are determined by $T$. This single-particle hopping process of the DW justifies the name ``domino''. Before $vT \gtrsim N$ (i.e., when the DW propagates to the right), $a_i$ is centered around $i=vT$ with a velocity $v$. Then, an initial state $\ket{+}_1\otimes |\bm{0}\rangle$ would evolve to \begin{equation}\label{eq:psido}
    \ket{\psi_{\rm do}}:=\ee^{-\ii TH_{\rm do}}\ket{+}_1\otimes |\bm{0}\rangle = \frac{1}{\sqrt{2}}\lr{ \ket{\bm{0}}+ \sum_{i=1}^N a_i \ket{1_i} },
\end{equation} 
which is roughly a GHZ-like state on the left $\sim vT$ qubits (with other qubits in state $\ket{0}$), useful for metrology~\cite{metro_domino21}.

\begin{figure}[t]
\centering
\includegraphics[width=.6\textwidth]{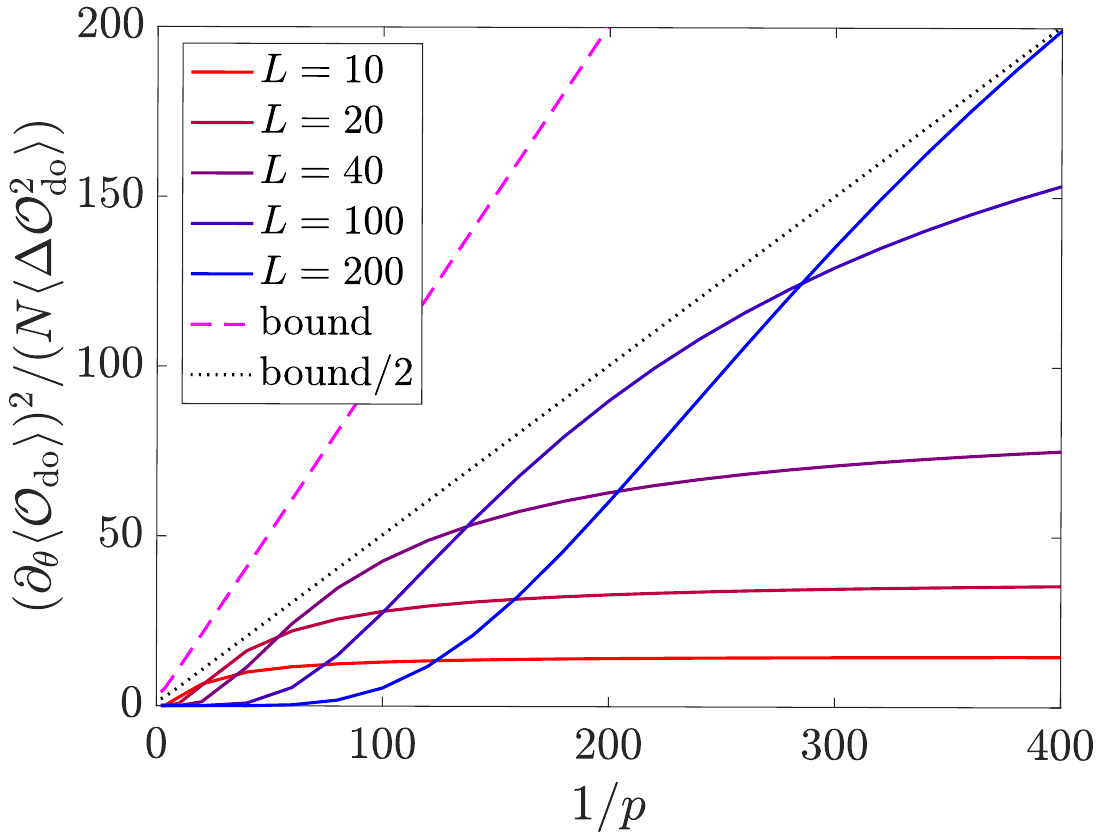}
    \caption{Optimal noisy metrology for $N=200$ qubits under dephasing, using partitioned domino state \eqref{eq:do_many+} generated from quantum domino dynamics, and on-site measurements $\OO_{\rm do}$ of the form \eqref{eq:do_manyO}. The $y$-axis quantifies $(\delta\theta)^{-2}/N$ for local estimation at $\theta = 0$. For each solid line, the spin chain is divided to $N/L$ groups each of size $L$, and the generating time $T\sim L$ is suitably chosen (see Appendix for details). For any noise probability $p\ll 1$ in \eqref{eq:Domega}, by choosing suitable $T$ and $L$, the performance approaches the bound in \eqref{eq:optimal} (pink dashed line) up to a factor $\approx 2$. }
    \label{fig:domino}
\end{figure}

Adapting the domino dynamics to our partition framework, we propose to evolve the initial state containing one $\ket{+}$ in every $L$ qubits: \begin{equation}\label{eq:do_many+}
    \ket{\Psi_{\rm do}} := \ee^{-\ii TH_{\rm do}}\lr{ \ket{+}_1\otimes\ket{+}_{L+1}\otimes \cdots} \otimes |\bm{0}\rangle_{\rm rest}.
\end{equation}
This final state, which we call the \emph{partitioned domino state}, satisfies \eqref{eq:C>LT} for $T\lesssim L$. 
We numerically study noisy metrology using $\ket{\Psi_{\rm do}}$ on a chain of $200$ qubits, and \figref{fig:domino} shows the results. Remarkably, this protocol is not only optimal asymptotically, but also \emph{quantitatively}. By adjusting $L$ for different $p$, the precision $(\delta\theta)^2$ is only roughly twice larger than the known lower bound \eqref{eq:optimal}. 

\revise{Interestingly, the QFI here even outperforms the quantum parity state \eqref{eq:GHZ_e} with a prefactor $1/2 > 1/\ee$. The quantum parity state, as a parallel metrology strategy, can be mapped to a sequential strategy, and this
 $\ee$-factor suppression of the QFI
is fundamental for any sequential metrology strategies \cite{demkowicz2014using}. 
Although the quantum parity state inspires our construction of the partitioned domino state, it turns out the latter contains entanglement structure that cannot be captured by mappings from sequential strategies. Intuitively, the partitioned domino state can be mapped to a ``coherent superposition'' of sequential strategies with varying duration for a quantum metrological register. The fact that such a superposition enhances metrological power could be interesting to investigate in future work.}

Besides having the optimal metrological power, the domino-state protocol requires only on-site measurements without any time-reversed evolution. To show this, we focus first on the simplest one-DW case with state \eqref{eq:psido} and prove the following in Appendix:

\begin{thm}\label{thm:localM}
In a rotated on-site basis (i.e., redefining states $\ket{1}_i$ with extra phases) such that $\mathrm{Arg}(a_i) =-\frac{\pi}{2}(1+i) -2\theta_{\rm pr} i$, define the measurement observable \begin{equation}\label{eq:O_do}
    \OO_{\rm do} := \sum_{vT/2\le i\le 2vT} |a_i| \fY_{1,i},
\end{equation} 
where $\fY_{1,i} := \bigotimes_{j=1}^i Y_j$.
If \begin{equation}\label{eq:a_i>c}
    \sum_{vT/2\le i\le 2vT} |a_i|^2 \ge c_{\rm do} \sim 1,
\end{equation}
then \begin{equation}\label{eq:Odo>T2}
    \frac{1}{\alr{\Delta \OO^2_{\rm do}}_{\rho_\theta}}\lr{\partial_\theta\alr{\OO_{\rm do}}_{\rho_\theta}}^2 \sim T^2,
\end{equation}
if noise $p\le c_{\rm do}/(vT)$ for a constant $c_{\rm do}$.
\end{thm}

\eqref{eq:O_do} can be realized by $Y$-basis 
on-site measurements with efficient classical post-processing.
The condition \eqref{eq:a_i>c} means that $|a_i|^2$, as a probability distribution, is concentrated around $vT$, which holds before $vT \gtrsim N$.

The idea to choose \eqref{eq:O_do} as the observable is as follows. $\ket{\psi_{\rm do}}$ is roughly a superposition of GHZ states $\ket{\GHZ}_{1\cdots i}\otimes \ket{\bm{0}}$ of different size $i$. For each GHZ state, $\fY_{1,i}$ is an optimal observable to sense the \(Z\)-axis magnetic field. However, each such observable does not have large sensitivity, i.e., $\partial_\theta \alr{\fY_{1,i}}\sim |a_i|i\ll i$, because the DW wave packet spreads. Instead, we use $\OO_{\rm do}$, a linear combination of them, to accumulate the small derivatives into $\partial_\theta \OO_{\rm do} \sim \sum_i |a_i|^2 \sim 1$. $\alr{\Delta \OO^2_{\rm do}}_{\rho_\theta}$ is also bounded by a constant due to the lack of interference among different GHZ states in $\ket{\psi_{\rm do}}$, making $\OO_{\rm do}$ a suitable observable. 

Generalizing the idea, the observable we use in \figref{fig:domino} for state \eqref{eq:do_many+} is of the form (see Appendix) 
\begin{equation}\label{eq:do_manyO}
    \sum_{\text{integer }I}\,\sum_{1-L/2\le i\le j\le L/2}b_{i,j}\fY_{IL+i,IL+j}~,
\end{equation}
where the real coefficients $b_{i,j}$ come from simulating the two-DW subspace for the group $I=0$,
which only requires \emph{polynomial} (in $L$) resource of classical computation. 

\section{Spin-squeezing}
We have established in 
\thmref{thm1} a sufficient condition for metrological optimality. Here we show this small-correlation condition (\eqref{eq:sumJCIJ<'}) is \emph{not necessary}: spin-squeezed states (SSS)~\cite{squeez_93,squeez_rev11} do not satisfy it in general, but can be optimal (see Appendix). 
\begin{thm}
\label{thm:squeeze}
    Suppose state $\rho$ satisfies \begin{align}
    \alr{X}_\rho &= \Theta(N), \quad
        \alr{\Delta Y^2}_\rho = N \xi^2, \label{eq:squeeze2} \\
        \alr{\Delta S^2}_\rho &= \order(N/\xi^2),\quad \forall S:=\textstyle\sum_i S_i \text{ with } \norm{S_i}=1 \label{eq:squeeze3}
    \end{align}
    where $X=\sum_i X_i$ etc and $S_i$ denotes an operator on $i$. Its metrological power is robust in noisy metrology with $p=\order(\xi^2)$: \begin{equation}\label{eq:noisy_sq}
        \mathcal{F}(\rho_\theta) \geq \frac{1}{\alr{\Delta Y^2}_{\rho_\theta}}\lr{\partial_\theta\alr{Y}_{\rho_\theta}}^2\Big|_{\theta=0} = \Theta( N/\xi^2).
    \end{equation}
\end{thm}
\thmref{eq:squeeze3} generalizes special SSSs and noise analyzed previously \cite{ulam2001spin,channel_estimate21,brask2015improved}, \revise{where it was shown that SSS could saturate the numerical prefactor in~\eqref{eq:optimal} as well}.  \eqref{eq:squeeze2} is a standard definition of SSS: $\rho$ is polarized in the $X$ direction and squeezed by a factor $\xi\ll 1$ in the $Y$ direction. For example, extremely squeezed states with $\xi=\Theta(1/\sqrt{N})$ exist in the permutation-symmetric subspace of the $N$ qubits, for which any two groups of qubits have a constant correlation $C_\rho(I;J)=\Theta(1)$, violating \eqref{eq:sumJCIJ<'}. Explanations on the violation of \eqref{eq:sumJCIJ<'} for any sufficiently small $\xi$ can be found in Appendix. 
Here we require the ``polarization'' $S$ along any direction in \eqref{eq:squeeze3} is at most stretched by a factor $\order(1/\xi)$, the minimum amount allowed by the uncertainty principle. The intuition is, under general noise, the observable $Y$ may mix with ``orthogonal'' directions $S$ and thus increase its variance. \eqref{eq:squeeze3} ensures its impact is sufficiently weak. The condition \eqref{eq:squeeze3} suggests the general usefulness of correlation bounds for noisy metrology. 

\revise{We remark that SSSs were known to achieve the optimal QFI (with the optimal prefactor) for parameter estimation in general noisy quantum channels, when concatenated with an inner layer of quantum error correcting codes~\cite{channel_estimate21,kobus2025asymptotically}. This endows SSSs with a particular advantage in noisy quantum metrology relative to other state families discussed previously, although achieving the optimal prefactor is not a primary focus of this work. }

In general, it is not straightforward to prepare SSSs from natural (i.e., time-independent) local dynamics: The squeezing parameter $\xi$ has a limited scaling with $N$ in recent proposals \cite{squeez_Norm23,squeeze_tower22,squeeze_2dlocal}, and it is argued that the 1d case can only generate $\xi\sim 1$ independent of $N$. 
This is in sharp contrast with the preparable domino state above, which has optimal robustness against noise and on-site measurement schemes. Nevertheless, it is unclear whether the domino dynamics can be generalized to higher dimensions. The small-correlation states and SSSs are thus complementary classes of metrological resources, with each offering advantages depending on the context.

\section{Discussion}
In this work, we prove sufficient conditions for optimal noisy metrology and illustrate them with several examples, from simple quantum parity states to quantum domino dynamics and spin-squeezed states.
Beyond state optimality, we also address the problem of efficient measurement schemes for metrology. In particular, the domino example suggests that on-site measurements apply to wider sensing scenarios beyond GHZ.

\vspace{2em}
\section*{Acknowledgements}
C.Y. was supported by the Department of Energy
under Quantum Pathfinder Grant DE-SC0024324. S.Z. acknowledges funding provided by Perimeter Institute for Theoretical Physics, a research institute supported in part by the Government of Canada through the Department of Innovation, Science and Economic Development Canada and by the Province of Ontario through the Ministry of Colleges and Universities.
V.V.A. thanks Ryhor Kandratsenia and Olga Albert for providing daycare support throughout this work.

\newpage

\begin{appendix}
\renewcommand{\thesubsection}{\thesection.\arabic{subsection}}
\renewcommand{\thesubsubsection}{\thesubsection.\arabic{subsubsection}}
\setcounter{thm}{0}

\begin{center}
    {\large \textbf{Appendix}}
\end{center}

For illustrative purpose, we mainly focus on sensing $Z$ magnetic field under dephasing noise $\DD_{\theta,p}$ (\eqref{eq:Domega} in the main text), and prove the three Theorems in this restricted setting first. In the last Section, we prove the full versions in Theorem \ref{suppthm:generalize}, after formal definitions of general noise. The reason for such arrangement is two-fold: First, the general case requires heavier notations and sometimes more complicated proof techniques. Second, dephasing is intuitively the most dangerous type of noise that would ruin sensing $Z$ field, so if a state is optimal against dephasing, it is expected to be optimal more generally.

We use a stronger non-asymptotic definition of big-O notation to avoid issues like which of the two limits $N\rightarrow \infty$ and $p\rightarrow 0$ are taken first. 
Namely, we define 
\begin{align*}
    f=\order(g)&\quad\quad\leftrightarrow\quad \quad\phantom{c_{2~}g\leq}f\leq c_{1}g\\f=\Theta(g)&\quad\quad\leftrightarrow\quad\quad c_{2}g\leq f\leq c_{3}g\\f=\Omega(g)&\quad\quad\leftrightarrow\quad\quad c_{4}g\leq f\\
\end{align*}
for some positive constants $c_{1,2,3,4}$. Here, $f$ and $g$ are any positive functions of $N$ (any integer $\geq 3$), $p$ (any real number $\in (1/N,1/2)$), $M$ (any integer satisfying \eqref{suppeq:p<G}), and all other parameters whose domains are specified in the contexts. In other words, we say $f=\order(g)$ if there exists a constant $c_{1,2,3,4}$ independent of $N,p,M$, such that $f\leq c_1 g$ holds for any parameters $N,p,\cdots$ in the aforementioned regime.

\section{Proof of Theorem 1: large QFI per group and small inter-group correlation guarantee optimal noisy metrology}

We rephrase Theorem 1 in the main text here. We consider here Pauli-Z signal and dephasing noise for simplicity, and the extension to general signal and noise models will be explained in Section~\ref{sec:general}. 
\begin{thm}\label{suppthm1}
    For each group $I$, suppose it has large QFI in the noiseless limit \begin{equation}\label{suppeq:FI0>G2}
        \cF\lr{\rho^{(0)}_{I,\theta}} \ge c^2 \G^2,
    \end{equation}
    and its correlation with other groups is bounded by 
    \begin{equation}
    \sum_{J: J\notin \mathcal{S}_I} C_\rho(I;J) \le c'\G^{-2}, \where 
    \abs{\mathcal{S}_I} \le c'', \label{suppeq:sumJCIJ<'}
\end{equation}
for some set $\mathcal{S}_I$.
Here, $\mathcal{S}_I\ni I$ contains a finite number of groups that may have large correlation with $I$. ($c,c',c''$ and $c_{\rm no}$ below can be any positive constants independent of $N,p$.)
Then for noise \begin{equation}\label{suppeq:p<G}
    p\le c_{\rm no}/\G,
\end{equation}
there exists an observable $\OO=\sum_I \OO_I$ such that \begin{equation}\label{suppeq:F>O>NG}
    \cF(\rho_\theta)\ge \frac{1}{\alr{\Delta \OO^2}_{\rho_\theta}}\lr{\partial_\theta\alr{\OO}_{\rho_\theta}}^2 =\Theta( N\G)
\end{equation}
\end{thm}


\begin{proof}
Our proof strategy is as follows:
Following the intuition of the partitioned GHZ example,
we first show that QFI of each group \eqref{suppeq:FI0>G2} is robust to noise \eqref{suppeq:p<G} such that 
\begin{equation}\label{suppeq:FI>G2}
    \cF\lr{\rho_{I,\theta};Z_I} = \Theta( \G^2),
\end{equation} 
where $Z_I:=\sum_{i\in I}Z_i$, and \begin{equation}
    \cF\lr{\sigma;A} := 2\sum_{\alpha,\beta: p_\alpha+p_\beta>0}\frac{(p_\alpha-p_\beta)^2}{p_\alpha+p_\beta} \abs{\alr{\alpha|A|\beta}}^2,
\end{equation}
given the eigendecomposition of $\sigma$: \begin{equation}\label{suppeq:sigma=palpha}
    \sigma = \sum_\alpha p_\alpha\ket{\alpha} \bra{\alpha}.
\end{equation}
Note that beyond the case when $\partial_\theta \rho_{I,\theta} = -\ii [Z_I,\rho_{I,\theta}]$, $\cF\lr{\rho_{I,\theta};Z_I}$ is not equal to $\cF\lr{\rho_{I,\theta}}$ in general.
We then show that \eqref{suppeq:FI>G2} guarantees the existence of an observable $\OO_I$ acting in the group that saturates \eqref{suppeq:FI>G2} parametrically. Furthermore, the $\OO_I$'s for different groups are roughly independent using the bound \eqref{suppeq:sumJCIJ<'} on inter-group correlation, so that \eqref{suppeq:F>O>NG} follows roughly by multiplying \eqref{suppeq:FI>G2} with the number of groups $N/\G$. 

More precisely, let $\DD_{0,p}$ be the pure dephasing channel on a single qubit. 
Since $\rho_{I,\theta}= \DD_{0,p}^{\otimes \G}(\rho^{(0)}_{I,\theta})$ and $\cF\lr{\rho^{(0)}_{I,\theta};Z_I}=\cF\lr{\rho^{(0)}_{I,\theta}}$, the continuity of QFI (see Eq.(A7) in \cite{QFI_contin16}; see also \cite{QFI_contin19}) implies \begin{align}\label{suppeq:FI-FI0}
    &\abs{\mathcal{F}(\rho_{I,\theta};Z_I) - \mathcal{F}(\rho_{I,\theta}^{(0)})} \le 32 \G^2 \sqrt{\norm{\DD_{0,p}^{\otimes \G}(\rho^{(0)}_{I,\theta}) - \rho^{(0)}_{I,\theta}}_1} \nonumber\\
    &\quad\le 32 \G^2 \sqrt{\norm{(\DD_{0,p}\otimes \id)(\rho^{(0)}_{I,\theta}) - \rho^{(0)}_{I,\theta}}_1 + \norm{(\DD_{0,p}\otimes \DD_{0,p} \otimes \id)(\rho^{(0)}_{I,\theta})) - (\DD_{0,p}\otimes \id)(\rho^{(0)}_{I,\theta}) }_1+\cdots} \nonumber\\
    &\quad \le 32 \G^2 \sqrt{2p\G} \le 32 \sqrt{2c_{\rm no}} \G^2,
\end{align}
where $\norm{\cdot}_1$ denotes the trace norm, $\id$ denotes the identity operation (or the identity operator as used later), the second line uses triangle inequality to give an upper bound containing $\G$ terms, and the third lines uses $\norm{\DD_{0,p}(\sigma)-\sigma}_1 = \norm{-p\sigma + p Z_i\sigma Z_i}_1\le 2p$ for any density matrix $\sigma$.
Then according to \eqref{suppeq:FI0>G2}, \eqref{suppeq:FI>G2} holds as long as $c_{\rm no}$ is a sufficiently small constant $32 \sqrt{2c_{\rm no}}\le c^2/2$.

Then we invoke the following Lemma that we prove later: 
\begin{lem}\label{lem:normO}
There exists a positive constant $c_{\rm tri}=\Theta(1)$ such that the following holds:
For any $2^\G$-dimensional state $\sigma_\theta$ whose dependence on $\theta$ is given by \begin{equation}\label{suppeq:dsigma=A}
    \partial_\theta \sigma_\theta = -\ii[A,\sigma_\theta]+ B,
\end{equation}
for some operators $A,B$ with $A$ normalized by $\norm{A}=M$, 
there exists an observable $\mathcal{L}$ such that \begin{subequations}\label{suppeq:L2eqs}
\begin{align}
    \partial_\theta \alr{\mathcal{L}}_{\sigma_\theta} &\ge \frac{1}{2\G}\mathcal{F}(\sigma_\theta;A)-c_{\rm tri}\G\norm{B}_1, \label{suppeq:dL>MB}\\
    \alr{\Delta \mathcal{L}^2}_{\sigma_\theta} &\le 1.
\end{align}
\end{subequations}
Furthermore, it has bounded operator norm \begin{equation}\label{suppeq:L<A}
    \norm{\mathcal{L}}\le c_{\rm tri}\G.
\end{equation}

\end{lem}

Then in the case when $\mathcal{F}(\sigma_\theta; A)=\Theta(\G^2)$ and $B$ is sufficiently small, measuring $\mathcal{L}$ achieves the HL while at the same time has not-too-large operator norm (e.g., not $\norm{\mathcal{L}}=\Theta( \norm{A} 2^\G)$).
Note that the observable $\mathcal{L}$ is similar to but not identical with the symmetric logarithmic derivative~\cite{estimation_tech09}: We define $\mathcal{L}$ in \eqref{suppeq:L=A} below, so that we can bound its operator norm. It is an interesting open question whether \eqref{suppeq:L<A} can be further improved (e.g., getting rid of the $\G$ dependence that allows $\G^{-2}$ to be removed from \eqref{suppeq:sumJCIJ<'}), and whether the operator norm of the symmetric logarithmic derivative can also be bounded.

In our case, let \begin{equation}\label{suppeq:sigma=rhoI}
    \sigma_\theta=\rho_{I,\theta},
\end{equation}
and the generator \begin{equation}\label{suppeq:A=ZI}
    A=Z_I.
\end{equation}
For the dephasing noise here, $B=0$ because the unitary evolution commutes with the noise channel. For more general noise in Section~\ref{sec:general}, we will show that \begin{equation}\label{suppeq:B<pM}
    \norm{B}_1 = \order(p\G).
\end{equation}
\eqref{suppeq:B<pM} implies that with a sufficiently small constant $c_{\rm no}$ in \eqref{suppeq:p<G}, the second term in \eqref{suppeq:dL>MB} $= \order(p\G^2)$ is subdominant comparing to the first $= \Theta(\G)$, so that Lemma~\ref{lem:normO} implies the existence of an observable $\OO_I$ that parametrically saturates the single-group QFI (\eqref{suppeq:FI>G2}): 
\begin{gather}
\label{suppeq:oi-1}
    \partial_\theta \alr{\OO_I}_{\rho_\theta} =\Theta(\G),\quad 
    \alr{\Delta \OO_I^2}_{\rho_\theta} \le 1,\\ 
\label{suppeq:oi-2}
    \norm{\OO_I} = \order(M), 
\end{gather}
leading to \begin{equation}\label{suppeq:dO=N_G}
    \partial_\theta \alr{\OO}_{\rho_\theta} =N/\G\cdot \Theta(\G) = \Theta(N),
\end{equation}
by summing over $I$.

Finally, to bound the variance, we derive a correlation bound for the final state $\rho_\theta$ from the initial one \eqref{suppeq:sumJCIJ<'}:
\begin{lem}\label{lem:cor_decrease}
    Suppose \eqref{suppeq:sumJCIJ<'} holds. Then for any set of operators $\{A_J\}$ normalized by \begin{equation}\label{suppeq:DeltaAJ<1}
        \alr{\Delta A_J^2}_{\rho_\theta}\le 1,
    \end{equation}
    we have \begin{equation}\label{suppeq:sumJnotI}
        \sum_{J\neq I}\abs{C_{\rho_\theta}(A_I, A_J) }\le c''-1+c'\G^2 \mathsf{A}^2,\;\;\forall I,\;\;\text{where}\;\;\mathsf{A}=\max_J \norm{A_J}.
    \end{equation}
\end{lem}
We will prove this Lemma in Section~\ref{sec:cor_decrease}. Note that this is the only place that uses \eqref{suppeq:sumJCIJ<'}. In other words, the condition \eqref{suppeq:sumJCIJ<'} can be replaced by a weaker condition \eqref{suppeq:sumJnotI}, and Theorem~\ref{suppthm1} still holds. Nevertheless, we use \eqref{suppeq:sumJCIJ<'} because it does not depend on the noise channel.

Let $A_I$'s in Lemma~\ref{lem:cor_decrease} be the observables $\OO_I$'s, we bound the total variance: \begin{align}\label{suppeq:DeltaO2<}
    \alr{\Delta \OO^2}_{\rho_\theta} &=\sum_I \alr{\Delta \OO_I^2}_{\rho_\theta} +\sum_I \sum_{J \neq I}C_{\rho_\theta}(\OO_I, \OO_J) \nonumber\\
    &\le N/\G \lr{1+ c''-1  + c'\G^{-2} \order(\G^2) } \nonumber\\
    &=\Theta(N/\G),
\end{align}
which combines with \eqref{suppeq:dO=N_G} to yield \eqref{suppeq:F>O>NG}. 

\end{proof}

\subsection{Proof of Lemma~\ref{lem:normO}: Optimal observable with bounded operator norm}
\begin{proof}

Using the eigendecomposition \eqref{suppeq:sigma=palpha} for $\sigma_\theta$ with $p_\alpha\ge 0$,
define a Hermitian observable as \begin{equation}\label{suppeq:L=A}
    \mathcal{L} = \frac{\ii}{M} \sum_{\alpha\beta} \mathrm{sign}(p_\alpha-p_\beta) \alr{\alpha|A|\beta} \ket{\alpha}\bra{\beta},
\end{equation}
where the sign function $\mathrm{sign}(p_\alpha-p_\beta)=1,-1,0$ if $p_\alpha-p_\beta> 0$, $<0$, $=0$ respectively. Assume $p_\alpha$ decreases, after arranging the basis $\{\ket{\alpha}\}$ in order. Up to a factor of $\ii/M$, $\mathcal{L}$ is equal to $A$ after the sign of the lower triangular part is flipped. Such operation does not change the norm of the operator too much. According to Eq.~(15) in~\cite{upper_triangle}, \begin{equation}
    \norm{\mathcal{L}} \le \frac{1}{\G} \norm{\text{upper triangular part of } A}\le \frac{1}{\G} \lr{\frac{8}{\pi^2}\log 2^\G + \mathrm{O}(1)} \norm{A},
\end{equation}
which leads to \eqref{suppeq:L<A} for some constant $c_{\rm tri}$.

We then verify \eqref{suppeq:L2eqs} by explicit computation, \begin{align}\label{suppeq:dthetaL=}
    \partial_\theta \alr{\mathcal{L}}_{\sigma_\theta} &= \frac{1}{\G}\sum_{\alpha\beta} \abs{p_\alpha-p_\beta} \abs{\alr{\alpha|A|\beta} }^2 + \tr{\cL B} \nonumber\\
    &\ge \frac{1}{\G}\sum_{\alpha\beta: p_\alpha+p_\beta>0}\frac{(p_\alpha-p_\beta)^2}{p_\alpha+p_\beta} \abs{\alr{\alpha|A|\beta}}^2  - \norm{B}_1 \norm{\cL} \nonumber\\
    &\ge \frac{1}{2\G}\mathcal{F}(\sigma_\theta;A)-c_{\rm tri}\G \norm{B}_1,
\end{align}
where we have used $|p_\alpha-p_\beta|<p_\alpha+p_\beta$ in the second line, and \eqref{suppeq:L<A} in the last line. Furthermore, \begin{equation}
    \alr{\Delta \mathcal{L}^2}_{\sigma_\theta}=\alr{\mathcal{L}^2}_{\sigma_\theta} \le \frac{1}{\G^2} \sum_{\alpha\beta} p_\alpha \abs{\alr{\alpha|A|\beta} }^2 \le \frac{1}{\G^2} \sum_{\alpha} p_\alpha \norm{A\ket{\alpha}}^2 \le \frac{1}{\G^2}\norm{A}^2\sum_\alpha p_\alpha =1,
\end{equation}
because $\sigma$ is normalized.

\end{proof}

\subsection{Proof of Lemma~\ref{lem:cor_decrease}: Correlations do not increase by local channels}\label{sec:cor_decrease}

\begin{proof}

For the groups in $\mathcal{S}_I$ that may have large correlation with $I$, we have
\begin{equation}\label{suppeq:Jin}
    \sum_{J\in \mathcal{S}_I: J\neq I} \abs{C_{\rho_\theta}(A_I, A_J)} \le \sum_{J\in \mathcal{S}_I: J\neq I} \sqrt{C_{\rho_\theta}(A_I, A_I) C_{\rho_\theta}(A_J, A_J)}\le \sum_{J\in \mathcal{S}_I: J\neq I}1=\abs{\mathcal{S}_I} -1 \le c''-1,
\end{equation}
from Cauchy-Schwarz inequality and \eqref{suppeq:DeltaAJ<1}.

For the other groups, we have \begin{align}\label{suppeq:Jnotin}
    \sum_{J\notin \mathcal{S}_I}\abs{C_{\rho_\theta}(\OO_I, \OO_J)} &= \sum_{J\notin \mathcal{S}_I} \abs{C_{\rho}(\widetilde{\OO}_I, \widetilde{\OO}_J) } \le\sum_{J\notin \mathcal{S}_I}C_\rho(I;J) \nonumber\\
    &\le c'\G^{-2}\lr{\max_J \norm{\widetilde{\OO}_J}}^2 \le c'\G^{-2}\lr{\max_J \norm{\OO_J}}^2 ,
\end{align}
where $\widetilde{\OO}_J = \DD_{\theta,p}(\OO_J)$ (the dual map of $\DD_{\theta,p}$ is itself), and the second line comes from \eqref{suppeq:sumJCIJ<'} and $\norm{\widetilde{\OO}_J}\le \norm{\OO_J}$ (operator norm is non-increasing under any dual map). 

Adding \eqref{suppeq:Jin} with \eqref{suppeq:Jnotin} leads to \eqref{suppeq:sumJnotI}.
\end{proof}

\section{Simplifying single-group QFI for a pure state via Wigner-Yanase correlation functions}


The single-group QFI, i.e., the QFI of the RDM in a single group (\eqref{suppeq:FI0>G2}) does not have an explicit expression in general. However, if the entire state $\rho_\theta$ is pure, we show that it is almost equivalent to certain correlation functions, which is more intuitive and easier to work with. The connection turns out to come from the quantity called Wigner-Yanase skew information~\cite{Wigner63}: \begin{equation}\label{suppeq:skew}
    \mathcal{I}(\rho, A) := \tr{\rho A^2} - \tr{\sqrt{\rho} A\sqrt{\rho}A},
\end{equation} 
defined for any state $\rho$ and operator $A$.

\begin{prop}\label{prop1}
    For a pure global state $\rho=\ket{\psi}\bra{\psi}$, consider bipartitioning the qubits into $I$ and $I^{\rm c}$ (complement of $I$). Then the QFI for group $I$ in the noiseless case is equivalent to its Wigner-Yanase skew information up to a factor between $1$ and $2$: \begin{equation}\label{suppeq:IF2I}
    \mathcal{I}(\rho_{I, \theta}^{(0)}, Z_I)\le \mathcal{F}(\rho_{I, \theta}^{(0)})\le 2\, \mathcal{I}(\rho_{I, \theta}^{(0)}, Z_I).
\end{equation}
    Moreover, the Wigner-Yanase skew information can be expressed as \begin{equation}\label{suppeq:I=COR}
    \mathcal{I}(\rho_{I, \theta}^{(0)}, Z_I) = C_\psi(Z_I, Z_I) - C_\psi(Z_I, A^{\rm c}),
\end{equation}
for some operator $A^{\rm c}$ acting on $I^{\rm c}$ with norm $\norm{A^{\rm c}}\le 2\G$.
\end{prop}

\begin{proof}
\eqref{suppeq:IF2I} is known from~\cite{skew<F}, given that the generator evolving $\rho_{I, \theta}^{(0)}$ is \eqref{suppeq:A=ZI}. As a result, we only need to show \eqref{suppeq:I=COR}.

Writing \begin{equation}\label{suppeq:schmidt}
    \ket{\psi} = \sum_{\alpha} \sqrt{p_\alpha} \ket{\alpha}_I \otimes \ket{\alpha}_{I^{\rm c}},
\end{equation} 
by Schmidt decomposition
, we have \begin{equation}\label{suppeq:rho=sumalpha}
    \rho_{I,\theta}^{(0)} = \ee^{-\ii \theta Z_I} \lr{\sum_\alpha p_\alpha \ket{\alpha}_I \bra{\alpha}_I} \ee^{\ii \theta Z_I}.
\end{equation}
Defining \begin{equation}\label{suppeq:tZ=Z}
    \tZ_I:=Z_I - \bra{\psi}Z_I\ket{\psi},
\end{equation}
we have
\begin{align}\label{suppeq:I>cL2}
    \mathcal{I}(\rho_{I, \theta}^{(0)}, Z_I) &= \mathcal{I}\lr{\rho_{I, \theta}^{(0)}, \tZ_I} = \mathcal{I}\lr{\sum_\alpha p_\alpha \ket{\alpha}_I \bra{\alpha}_I, \tZ_I} \nonumber\\
    &= \sum_\alpha p_\alpha \bra{\alpha}\tZ_I^2 \ket{\alpha}_I - \sum_{\alpha \beta} \sqrt{p_\alpha p_\beta} \bra{\alpha}\tZ_I \ket{\beta}_I \bra{\alpha}\tZ_I \ket{\beta}^*_I \nonumber\\
    &= \sum_\alpha p_\alpha \bra{\alpha}\tZ_I^2 \ket{\alpha}_I - \sum_{\alpha \beta} \sqrt{p_\alpha p_\beta} \bra{\alpha}\tZ_I \ket{\beta}_I \bra{\alpha}A^{\rm c} \ket{\beta}_{I^{\rm c}}, \nonumber\\
    &=\bra{\psi}\tZ_I^2 \ket{\psi} - \bra{\psi} \tZ_I A^{\rm c}\ket{\psi}= C_\psi(Z_I, Z_I) - C_\psi(Z_I, A^{\rm c}),
\end{align}
where $^*$ denotes complex conjugate. 
Here the first line comes from the fact that the skew information \eqref{suppeq:skew} is unchanged when adding identity to the operator $Z_I$, and that the evolution $\ee^{-\ii \theta Z_I}$ commutes through $\tZ_I$. The second line comes from directly expanding \eqref{suppeq:skew}. In the third line, we defined an operator $A^{\rm c}$ acting on $I^{\rm c}$ by \begin{equation} \label{suppeq:A=ZI1}
    \bra{\alpha}A^{\rm c} \ket{\beta}_{I^{\rm c}} = \bra{\alpha}\tZ_I \ket{\beta}^*_I, 
\end{equation}
and setting other matrix elements to $0$. 
The last line of \eqref{suppeq:I>cL2} simplifies the terms by invoking the Schmidt decomposition \eqref{suppeq:schmidt}, and the definition of the connected correlation function noticing $\bra{\psi} \tZ_I \ket{\psi} = 0$.

Finally,
\begin{equation}
    \norm{A^{\rm c}} =\norm{(P \tZ_I P)^*} = \norm{P \tZ_I P} \le \norm{\tZ_I} \le 2\norm{Z_I}= 2M.
\end{equation}
from \eqref{suppeq:A=ZI1}, where $P$ is the projector onto $\mathrm{span}\{\ket{\alpha}_I,\forall \alpha\}$.
\end{proof}

\subsection{Application: Large correlation for locally generated states implies large QFI per group}
Here we show an application of Proposition~\ref{prop1}: In the main text, the condition \eqref{eq:C>LT} for Theorem 2 implies condition \eqref{eq:FI0>G2} in Theorem 1, so the locally generated states for Theorem 2 are special cases of states in Theorem 1. Note that Theorem 2 holds by itself (as we will prove in the next Section) independent of discussion here.

To show \eqref{eq:C>LT} implies \eqref{eq:FI0>G2} in the main text, we only need to show \begin{equation}\label{suppeq:I=L2d}
    \mathcal{I}(\rho_{I, \theta}^{(0)}, Z_I) = C_\psi(Z_I, Z_I) - C_\psi(Z_I, A^{\rm c}) = \Theta(L^{2d}),
\end{equation}
according to Proposition~\ref{prop1}, with $A^{\rm c}$ defined in \eqref{suppeq:A=ZI1}. Since the first term $C_\psi(Z_I, Z_I)=\Theta(L^d T^d)$, \eqref{suppeq:I=L2d} will hold as long as the second term scales as $\order(L^{d-1}T^{d+1})$, and $L/T$ is chosen as a sufficiently large constant. To bound the second term $C_\psi(Z_I, A^{\rm c})$, we separate \begin{equation}
    Z_I = Z_{\rm bulk} + Z_{\rm boundary}, \where Z_{\rm boundary}=\sum_{i\in I: \mathsf{d}(i,I^{\rm c})\le 3T} Z_i,
\end{equation}
(and $\widetilde{Z}_{\rm bulk}$, $\widetilde{Z}_{\rm boundary}$ similarly by \eqref{suppeq:tZ=Z}; here the index $I$ is omitted for simplicity). 
Here $Z_{\rm boundary}$ ($Z_{\rm bulk}$) are summed over spins $i\in I$ close to (far from) the boundary. Since the correlation length is roughly $2T$, \begin{equation}\label{suppeq:ZbulkA=0}
    C_\psi(Z_{\rm bulk}, A^{\rm c})\approx 0,
\end{equation}
(see the later rigorous \eqref{suppeq:LRB_cor}). On the other hand, since \eqref{suppeq:A=ZI1} is linear, we can also separate \begin{equation}
    A^{\rm c} = A^{\rm c}_{\rm bulk} + A^{\rm c}_{\rm boundary}, \where  \bra{\alpha}A^{\rm c}_{\rm boundary} \ket{\beta}_{I^{\rm c}} = \bra{\alpha}\tZ_{\rm boundary} \ket{\beta}^*_I
\end{equation}
accordingly. The contribution from $A^{\rm c}_{\rm bulk}$ is then also negligible by $Z\leftrightarrow A$ symmetry: 
\begin{align}\label{suppeq:ZAbulk=0}
    C_\psi(Z_I, A^{\rm c}_{\rm bulk}) &=\sum_{\alpha \beta} \sqrt{p_\alpha p_\beta} \bra{\alpha}\tZ_I \ket{\beta}_I \bra{\alpha}A^{\rm c}_{\rm bulk} \ket{\beta}_{I^{\rm c}} \nonumber\\
    &= \sum_{\alpha \beta} \sqrt{p_\alpha p_\beta} \bra{\alpha}\tZ_I \ket{\beta}_I \bra{\alpha}Z_{\rm bulk} \ket{\beta}_{I}^* \nonumber\\
    &= \lr{\sum_{\alpha \beta} \sqrt{p_\alpha p_\beta} \bra{\alpha}\tZ_I \ket{\beta}_I^* \bra{\alpha}Z_{\rm bulk} \ket{\beta}_{I}  }^*\nonumber\\
    &= \lr{\sum_{\alpha \beta} \sqrt{p_\alpha p_\beta} \bra{\alpha}A^{\rm c} \ket{\beta}_I \bra{\alpha}Z_{\rm bulk} \ket{\beta}_{I}  }^* = C_\psi(Z_{\rm bulk}, A^{\rm c})^* \approx 0.
\end{align}
As a result, \begin{align}\label{suppeq:CZA=}
    C_\psi(Z_I, A^{\rm c}) &= C_\psi(Z_I, A^{\rm c}_{\rm bulk})+ C_\psi(Z_{\rm bulk}, A^{\rm c}_{\rm boundary})+ C_\psi(Z_{\rm boundary}, A^{\rm c}_{\rm boundary}) \nonumber\\
    &\approx 0+0+ C_\psi(Z_{\rm boundary}, A^{\rm c}_{\rm boundary}) \nonumber\\
    &= \tr{\sqrt{\rho} \tZ_{\rm boundary}\sqrt{\rho}\tZ_{\rm boundary}} \le \tr{\rho \tZ_{\rm boundary}^2} = C_\psi(Z_{\rm boundary}, Z_{\rm boundary})\nonumber\\
    &=\order(L^{d-1}T\times T^d)=\order(L^{d-1}T^{d+1}).
\end{align}
Here in the second line, we have used \eqref{suppeq:ZAbulk=0} and \eqref{suppeq:ZbulkA=0} where $A^{\rm c}$ is replaced by $A^{\rm c}_{\rm boundary}$ (because the only crucial thing is that $Z_{\rm bulk}$ is far from $I^{\rm c}$). In the third line, we have used Cauchy-Schwarz inequality \begin{equation}
    \tr{\sqrt{\rho} \OO\sqrt{\rho} \OO} = \tr{ \lr{\OO\sqrt{\rho}}^\dagger\sqrt{\rho} \OO}\le \sqrt{\tr{ \lr{\OO\sqrt{\rho}}^\dagger\OO\sqrt{\rho}} \tr{ \lr{\sqrt{\rho} \OO}^\dagger\sqrt{\rho} \OO}} =\tr{\rho \OO^2} ,
\end{equation}
for Hermitian operator $\OO=\tZ_{\rm boundary}$. The last line of \eqref{suppeq:CZA=} uses the fact that each $i$ is correlated to $\order(T^d)$ other qubits, while there are $\order(L^{d-1}T)$ qubits included in $Z_{\rm boundary}$.

To conclude, we have shown that the second term of \eqref{suppeq:I=L2d} is subdominant comparing to the first term. Thus, \eqref{eq:C>LT} implies \eqref{eq:FI0>G2} in the main text.

\section{Proof of Theorem 2: on-site measurement scheme with time-reversal}\label{sec:thm2}
\subsection{Lieb-Robinson bound: Rigorous treatment of light cones}
We first introduce the rigorous version of \eqref{eq:AT=} in the main text. Suppose the Hamiltonian evolution time during $U$ is actually $T'$ instead of $T$ (although we still use notation $A(T)=U^\dagger A U$), the following Lieb-Robinson bound~\cite{Lieb1972,review_us23} holds: There exist constants $\mu,\mu',v_{\rm LR}=\Theta(1)$, such that for any operator $A$ supported on set $S\subset\Lambda$, there exists an operator $\widetilde{A}_{\widetilde{S}}$ acting on $\widetilde{S}=\{i\in \Lambda: \mathsf{d}(i,S)\le T\}$ such that \begin{equation}\label{suppeq:LRB}
    \norm{A(T)-\widetilde{A}_{\widetilde{S}}}\le \mu' \norm{A} \ee^{-\mu (T-v_{\rm LR} T')}.
\end{equation}
Therefore, we can choose \begin{equation}\label{suppeq:T=T'}
    T=v_{\rm LR}T'+\Theta(\log L),
\end{equation} such that \begin{equation}\label{suppeq:AT=}
    \norm{A(T)-\widetilde{A}_{\widetilde{S}}}\le \norm{A}\epsilon, \where \epsilon := L^{-\alpha},
\end{equation} 
for some constant $\alpha>0$. Later, we will see that $\epsilon$ always contributes in a subdominant way. The overhead $\log L \ll L = \Theta(T)$ in \eqref{suppeq:T=T'} is negligible so that we can still assume (\eqref{eq:C>LT} in the main text) \begin{equation}\label{suppeq:C>LT}
    C_\psi(Z_I, Z_I) \ge c_Z L^d T^d,
\end{equation}
which is equivalent to $C_\psi(Z_I, Z_I) = \Theta[L^d (T')^d]$.

As a consequence of the Lieb-Robinson bound \eqref{suppeq:LRB}, faraway regions have exponentially decaying correlation in $\psi$: \begin{equation}\label{suppeq:LRB_cor}
    \abs{C_\psi(A_I, B_J)} \le c_{\rm LRB}' \norm{A_I}\norm{B_J} \ee^{-c_{\rm LRB}\, \mathsf{d}(I,J)}, \quad \mathsf{d}(I,J) \ge 3v_{\rm LR} T',
\end{equation}
for some constants $c_{\rm LRB},c_{\rm LRB}'$, where $\mathsf{d}(I,J)=\min_{i\in I, j\in J}\mathsf{d}(i,j)$ is the distance between the two sets. See Theorem~5.9 in \cite{review_us23} for a proof, where a prefactor $|\partial I|=\Theta( L^{d-1})$ is absorbed into the exponential using $L=\Theta(T')$. As a result, \eqref{suppeq:sumJCIJ<'} is naturally satisfied for $\psi$, because apart from a finite number of neighboring $J$'s, \begin{equation}\label{suppeq:LRBcor}
    \sum_{J: \mathsf{d}(I,J) \ge 3T} \abs{C_\rho(A_I, A_J)} \le c_{\rm LRB}'\mathsf{A}^2 \sum_{J: \mathsf{d}(I,J) \ge 3T}\ee^{-c_{\rm LRB}\, \mathsf{d}(I,J)} \le \tilde{c}_{\rm LRB}\mathsf{A}^2 \ee^{-3c_{\rm LRB}T} \ll T^{-2d}\mathsf{A}^2.
\end{equation}
Here we have used that \eqref{suppeq:LRB_cor} also holds for $\mathsf{d}(I,J)\ge 3T$, and $\tilde{c}_{\rm LRB}$ is a constant from the convergent sum of the exponentially decaying terms.

\subsection{Proof of Theorem 2}
We repeat Theorem 2 in the main text for convenience:
\begin{thm}\label{suppthm:Orev>}
    If \eqref{suppeq:C>LT} holds, then
    there exist constants $c_{\rm no}$ and $C>c>1$ determined by $c_Z,d$, such that
    for any $C\ge L/T\ge c$ and mild noise \eqref{suppeq:p<G}, \begin{equation}\label{suppeq:O>NL}
    \frac{1}{\alr{\Delta \OO^2_{\rm rev}}_{\rho_\theta}}\lr{\partial_\theta\alr{\OO_{\rm rev}}_{\rho_\theta}}^2 = \Theta( N L^d).
\end{equation}
\end{thm}

\begin{proof}
We focus on $d=2$ for concreteness as shown in Fig.~2 of the main text, and other dimensions follow analogously. We first consider the noiseless case and invoke the following Lemma that we prove later in the next subsection.

\begin{lem}\label{lem}
For sufficiently large constant $c\le L/T$,
\begin{align}\label{suppeq:P=1-c2}
    -\partial_\theta \alr{\widetilde{P}_I}_{\rho_\theta^{(0)}} &= c_1 \tilde{\theta} L^4 + \order(\tilde{\theta}^2 L^6) \\
    &=\Theta(L^2), \quad \forall \tilde{\theta}\in [ \tilde{c}/(2L^2), \tilde{c}/L^2], \label{suppeq:-dtheta=L2}
\end{align}
for some constants $c_1,\tilde{c}=\Theta(1)$.
\end{lem}

According to \eqref{suppeq:-dtheta=L2}, $\abs{\partial_\theta\alr{\OO_{\rm rev}}_{\rho_\theta^{(0)}}}=\sum_I \Theta(L^2)=N$. On the other hand, \begin{align}\label{suppeq:DeltaO<}
    \alr{\Delta \OO^2_{\rm rev}}_{\rho_\theta^{(0)}} &=\sum_I \sum_J C_{U_{\rm rev} \rho_\theta^{(0)} U_{\rm rev}^\dagger}(P_I, P_J) \nonumber\\
    &\le \sum_I \order(1) = \order(N/L^2),
\end{align} 
which leads to \eqref{suppeq:O>NL} at $p=0$. \eqref{suppeq:DeltaO<} holds because $U_{\rm rev} \rho_\theta U_{\rm rev}^\dagger$ is short-range entangled, so only nearby groups $(I,J)$ have $C_{U_{\rm rev} \rho_\theta U_{\rm rev}^\dagger}(P_I, P_J) = \order(\norm{P_I}\norm{P_J}) = \order(1)$; the correlation between faraway groups are exponentially small and safely ignored.

Finally, we consider the noisy case $p>0$. \eqref{suppeq:DeltaO<} still holds when we replace the noiseless $\rho_\theta^{(0)}$ with the noisy $\rho_\theta$ because adding noise does not create long-range correlation~\cite{LRB_open10}. Therefore, we only need to show that \eqref{suppeq:-dtheta=L2} is robust. Viewing the dephasing channel as (its dual map) acting on the observable, \begin{align}\label{suppeq:P-P0<}
    &\abs{ \partial_\theta \lr{\alr{\widetilde{P}_I}_{\rho_\theta}- \alr{\widetilde{P}_I}_{\rho_\theta^{(0)}}} }= \abs{ \alr{[Z, \DD_{0,p}^{\otimes N}(\widetilde{P}_I\otimes \id)-\widetilde{P}_I\otimes \id ]}_{\rho_\theta^{(0)}} } \nonumber\\
    &\le 2 \norm{Z_{\widetilde{I}_T}} \norm{(\DD_{0,p}^{\otimes M'}\otimes \id)(\widetilde{P}_I\otimes \id)-\widetilde{P}_I\otimes \id} +\order(L^2\epsilon)\nonumber\\ 
    &\le 2 M'\sum_{g=1}^{M'}\norm{(\DD_{0,p}^{\otimes g}\otimes \id)(\widetilde{P}_I\otimes \id)-(\DD_{0,p}^{\otimes g-1}\otimes \id)(\widetilde{P}_I\otimes \id)}+\order(L^2\epsilon) \le \Theta(L^4) p+\order(L^2\epsilon).
\end{align}
Here in the second line, we have again used the locality of $\widetilde{P}_I$ to restrict $Z$ and the dephasing channel to region $\widetilde{I}_T$ defined by 
$\widetilde{I}_T=\{i\in \Lambda: \mathsf{d}(i,{\widetilde{I}})\le T\}$ with size $M'=\Theta(L^2)$. The $\order(L^2\epsilon)$ error comes from the exponentially decaying Lieb-Robinson bound \eqref{suppeq:LRB}, so that only $\Theta(L^2)$ many $Z_i$'s near $\widetilde{I}_T$ dominates. In the last line, we have used the triangle inequality, $\norm{\DD_{0,p}(A)-A}=\norm{p(Z_i A Z_i - A)}\le 2p\norm{A}$, and $\norm{(\DD_{0,p}^{\otimes g-1}\otimes \id)(\widetilde{P}_I\otimes \id)}\le \norm{P_I}\le 1$. Therefore, there exists constant $c_{\rm no}$ such that \eqref{suppeq:P-P0<} is smaller than \eqref{suppeq:-dtheta=L2} at weak noise \eqref{suppeq:p<G}, which further leads to \eqref{suppeq:O>NL} by summing over $I$ and combining with \eqref{suppeq:DeltaO<}.
\end{proof}

\subsection{Proof of Lemma~\ref{lem}: The local version of the Loschmidt echo}
\begin{proof}[Proof of Lemma~\ref{lem}]
Since the second line \eqref{suppeq:-dtheta=L2} follows immediately from the first line, we focus on showing the first line \eqref{suppeq:P=1-c2} for sufficiently large $c\le L/T$.
From direct calculation and Taylor's theorem, \begin{align}\label{suppeq:PI=expand}
    -\partial_\theta\alr{\widetilde{P}_I}_{\rho_\theta^{(0)}} &=-\partial_{\tilde{\theta}}\bra{\bm{0}} \ee^{\ii \tilde{\theta} Z(T)} P_{\widetilde{I}} \ee^{-\ii \tilde{\theta} Z(T)} \ket{\bm{0}} = -\ii \bra{\bm{0}} \ee^{\ii \tilde{\theta} Z(T)} [Z(T), P_{\widetilde{I}}] \ee^{-\ii \tilde{\theta} Z(T)} \ket{\bm{0}} \nonumber\\
    &= \tilde{\theta} \bra{\bm{0}}[Z(T), [Z(T), P_{\widetilde{I}}] ]\ket{\bm{0}} + \frac{\ii}{2}\tilde{\theta}^2 \bra{\bm{0}}\ee^{\ii \tilde{\theta}' Z(T)} [Z(T),[Z(T), [Z(T),  P_{\widetilde{I}} ] ]]\ee^{-\ii \tilde{\theta}' Z(T)}\ket{\bm{0}},
\end{align}
where $\tilde{\theta}'\in [0,\tilde{\theta}]$, and we have used that the zeroth order vanishes: \begin{equation} \label{suppeq:ZP=0}
    \bra{\bm{0}}[Z(T), P_{\widetilde{I}}] \ket{\bm{0}} = \bra{\bm{0}}Z(T)\ket{\bm{0}}-\bra{\bm{0}}Z(T)\ket{\bm{0}}=0.
\end{equation}
To bound the first term in \eqref{suppeq:PI=expand}, we expand $Z=\sum_J Z_J$ and separate it to two kinds of terms (ignoring the prefactor $\tilde{\theta}$): 

(\emph{i}) The single term \begin{align}\label{suppeq:ZIZI>}
    \bra{\bm{0}}[Z_I(T), [Z_I(T), P_{\widetilde{I}}] ]\ket{\bm{0}} &= 2\bra{\bm{0}}Z_I^2(T) P_{\widetilde{I}}\ket{\bm{0}} - 2\bra{\bm{0}}Z_I(T) P_{\widetilde{I}}Z_I(T) \ket{\bm{0}} \nonumber\\
    &= 2\bra{\bm{0}}Z_I^2(T) P_{\widetilde{I}}\ket{\bm{0}} - 2\bra{\bm{0}}Z_I(T) P_{\widetilde{I}}Z_I(T) P_{\widetilde{I}^{\rm c}}\ket{\bm{0}} \nonumber\\
    &= 2\bra{\bm{0}}Z_I^2(T) \ket{\bm{0}} - 2\bra{\bm{0}}Z_I(T) \lr{ P_{\widetilde{I}}\otimes P_{\widetilde{I}^{\rm c}}} Z_I(T) \ket{\bm{0}} + \order(L^4\epsilon)\nonumber\\
    &= 2\, C_\psi(Z_I, Z_I) + \order(L^4\epsilon) \ge 2c_Z L^2T^2+ \order(L^4\epsilon).
\end{align}
Here we have used $\ket{\bm{0}}=P_{\widetilde{I}^{\rm c}}\ket{\bm{0}}$ ($\rm c$ means complement) in the second line, and $\norm{[Z_I(T), P_{\widetilde{I}^{\rm c}} ]}\le \norm{Z_I}\epsilon=L^2\epsilon$ from \eqref{suppeq:AT=} in the third line with \begin{equation}
    \bra{\bm{0}}Z_I(T) P_{\widetilde{I}}Z_I(T) P_{\widetilde{I}^{\rm c}}\ket{\bm{0}} \le \norm{Z_I(T)} \norm{P_{\widetilde{I}}} \norm{[Z_I(T), P_{\widetilde{I}^{\rm c}} ]} \le L^2\cdot 1\cdot \epsilon.
\end{equation}
We will use such manipulations many times later.
In the last line of \eqref{suppeq:ZIZI>}, we have used \eqref{suppeq:C>LT}. One can see the $\epsilon$ term is subdominant because $\epsilon\ll 1$.

(\emph{ii}) All other terms can be arranged by Jacobi identity and $[Z_J(T), Z_K(T)]=0$ to the form \begin{equation}\label{suppeq:ZJZKP}
    \bra{\bm{0}}[Z_J(T), [Z_K(T), P_{\widetilde{I}}] ]\ket{\bm{0}},
\end{equation}  
where $K\neq I$ ($J$ can be $I,K$ or neither). In other words, if the original $K=I$, then $J\neq I$ (otherwise we return to the single term \eqref{suppeq:ZIZI>}), and we use \begin{equation}
    [Z_J(T), [Z_K(T), P_{\widetilde{I}}] = -[Z_K(T), [P_{\widetilde{I}},Z_J(T)]] - [P_{\widetilde{I}}, [Z_J(T), Z_K(T) ] ]=[Z_K(T), [Z_J(T), P_{\widetilde{I}}]],
\end{equation}
before relabeling $J\leftrightarrow K$. Moreover, we can assume $J,K$ in \eqref{suppeq:ZJZKP} are neighbors of $I$ ($J$ can also be $I$ itself), i.e. one of the $9$ blocks in Fig.~2 of the main text; otherwise for example $\norm{[Z_K(T), P_{\widetilde{I}}]}= \order(\ee^{-\mu \mathsf{d}(K,I)}) = \order(\ee^{-\mu L})$ is exponentially small from Lieb-Robinson bound \eqref{suppeq:LRB}, whose contribution to \eqref{suppeq:PI=expand} is negligible.

For \eqref{suppeq:ZJZKP}, since $K\neq I$, $Z_k(T)$ ($k\in K$) contributes only when it has support on $\widetilde{I}$, the red region in Fig.~2 of the main text. This requires $k\in \widetilde{I}_T=\{i\in \Lambda: \mathsf{d}(i,\widetilde{I})\le T\}$, otherwise $\norm{[Z_k(T), P_{\widetilde{I}}]}\le \epsilon$. This leads to the first line of \begin{align}\label{suppeq:ZJZKP=}
    \bra{\bm{0}}[Z_J(T), [Z_K(T), P_{\widetilde{I}}] ]\ket{\bm{0}} 
    &=\sum_{k\in K\cap \widetilde{I}_T}\bra{\bm{0}}[Z_J(T), [Z_k(T), P_{\widetilde{I}}] ]\ket{\bm{0}} + \order(L^4 \epsilon) \nonumber\\
    &= \sum_{k\in K\cap \widetilde{I}_T}\sum_{j\in J}\bra{\bm{0}}[Z_j(T), [Z_k(T), P_{\widetilde{I}}] ]\ket{\bm{0}} + \order(L^4 \epsilon) \nonumber\\
    &= \sum_{k\in K\cap \widetilde{I}_T}\sum_{j\in J: \mathsf{d}(j,k)\le 2T}\bra{\bm{0}}[Z_j(T), [Z_k(T), P_{\widetilde{I}}] ]\ket{\bm{0}} + \order(L^4 \epsilon)\nonumber\\
    &= \order(LT\cdot T^2)+ \order(L^4 \epsilon) = \order(LT^3).
\end{align}
Here the second line comes from expanding $Z_J=\sum_{j\in J} Z_J$. To get the third line of \eqref{suppeq:ZJZKP=}, observe that for any pairs of $j,k$ with $\mathsf{d}(j,k)>2T$, there exists bipartition $\widetilde{I}=\widetilde{I}_j\otimes \widetilde{I}_k$ such that $Z_j(T)$ roughly has no support on $\widetilde{I}_k$: $\norm{[Z_j(T), P_{\widetilde{I}_k}]}\le \epsilon$, and vice versa. Therefore, \begin{align}
    \bra{\bm{0}}[Z_j(T), [Z_k(T), P_{\widetilde{I}}] ]\ket{\bm{0}} &= \bra{\bm{0}}[Z_j(T), [Z_k(T), P_{\widetilde{I}_k}] \otimes P_{\widetilde{I}_j} + \order(\epsilon) ]\ket{\bm{0}} \nonumber\\
    &= \bra{\bm{0}}[Z_j(T),P_{\widetilde{I}_j}]\otimes [Z_k(T), P_{\widetilde{I}_k}] \ket{\bm{0}} + \order(\epsilon) \nonumber\\
    &= \bra{\bm{0}}[Z_j(T),P_{\widetilde{I}_j}]\ket{\bm{0}} \bra{\bm{0}}[Z_k(T), P_{\widetilde{I}_k}] \ket{\bm{0}} + \order(\epsilon) = \order(\epsilon),
\end{align}
where we have used \eqref{suppeq:ZP=0} at last. This justifies the third line of \eqref{suppeq:ZJZKP=}, which leads to the last line because the number of $j$ for each $k$ is bounded by $\order(T^2)$, while the number of $k$ is bounded by $\abs{K\cap \widetilde{I}_T}=\order(LT)$.

Since there are finitely many pairs of $(J,K)$ to choose, type (\emph{ii}) terms contribute $\order(LT^3)$. Combining with \eqref{suppeq:ZIZI>}, the first term in \eqref{suppeq:PI=expand} is $\tilde{\theta}\mlr{2c_Z L^2 T^2 - \order\lr{L^3 T}}=c_1 \tilde{\theta} L^4$ with $c_1=\Theta(1)$ for sufficiently large $L/T$. Namely, the single term of type (\emph{i}) dominates.

It remains to bound the second-order remainder term in \eqref{suppeq:PI=expand}. Using Jacobi identities and \eqref{suppeq:AT=} as above, the three $Z(T)$'s in the commutator can all be replaced by $Z_{\widetilde{I}_T}(T) + \order(L^2\epsilon)$. So the second term in \eqref{suppeq:AT=} is bounded by \begin{equation}
    \order\lr{\tilde{\theta}^3 \lVert{Z_{\widetilde{I}_T}\rVert}^3 } + \order(L^6\epsilon) = \order\mlr{\tilde{\theta}^3 (L+2T)^6}=\order\lr{\tilde{\theta}^3L^6}.
\end{equation}
\end{proof}

\section{States generated from quantum domino dynamics}

\subsection{Proof of Theorem 3: On-site measurements for quantum domino dynamics}
We restate Theorem 3 in the main text here:
\begin{thm}\label{suppthm:localM}
Consider the single-DW state $\ket{\psi_{\rm do}}$ in \eqref{eq:ZIZI>} in the main text.
In a rotated on-site basis (i.e. redefining the local states $\ket{1}_i$ with an extra phase) such that $\mathrm{Arg}(a_i) =-\frac{\pi}{2}(1+i) -2\theta_{\rm pr} i$, define the measurement observable \begin{equation}\label{suppeq:O_do}
    \OO_{\rm do} = \sum_{vT/2\le i\le 2vT} |a_i| \fY_{1,i},
\end{equation} 
where $\fY_{1,i} := \bigotimes_{j=1}^i Y_j$.
For any $v,T > 0$, if
\begin{equation}\label{suppeq:a_i>c} 
    \sum_{vT/2\le i\le 2vT} |a_i|^2 \ge c_{\rm do} = \Theta(1),
\end{equation}
then \begin{equation}\label{suppeq:Odo>T2}
    \frac{1}{\alr{\Delta \OO^2_{\rm do}}_{\rho_\theta}}\lr{\partial_\theta\alr{\OO_{\rm do}}_{\rho_\theta}}^2 = \Theta(T^2),
\end{equation}
for $\rho_\theta=\DD_{\theta,p}^{\otimes N}(\ket{\psi_{\rm do}}\bra{\psi_{\rm do}})$
under noise $p\le 1/(vT)$. 
\end{thm}

\begin{proof}

We first focus on the noiseless case. Direct computation yields expectation values \begin{equation}\label{suppeq:fY=0}
    \alr{\fY_{1,i}}_{\rho_\theta^{(0)}} = 0, \quad \partial_\theta\alr{\fY_{1,i}}_{\rho_\theta^{(0)}} = 2|a_i|i,
\end{equation}
where we have used the redefined local states $\ket{1}_i$ to make sure all $\partial_\theta\alr{\fY_{1,i}}_{\rho_\theta^{(0)}}$ are at its maximum. Note that $i$ is the real number for qubit index, not the imaginary unit $\ii$.
Although each $\fY_{1,i}$ does not have HL sensitivity because $|a_i|\ll 1$ and $\alr{\Delta \fY_{1,i}^2}_{\rho_\theta^{(0)}}=\alr{\fY_{1,i}^2}_{\rho_\theta^{(0)}}-0=1$, a linear combination of them like \eqref{suppeq:O_do} does. Here we assume a more general form $\OO_{\rm do} = \sum_{i=1}^N b_i \fY_{1,i}$ for latter convenience.
\eqref{suppeq:fY=0} then yields $\alr{\OO_{\rm do}}_{\rho_\theta^{(0)}}=0$ and \begin{equation}\label{suppeq:dphi>T}
    \partial_\theta\alr{\OO_{\rm do}}_{\rho_\theta^{(0)}} = 2\sum_i b_i|a_i|i \ge vT \sum_{vT/2\le i\le 2vT} b_i|a_i|.
\end{equation}
We then want to show the variance of $\OO_{\rm do}$ is small in the state: \begin{align}\label{suppeq:Odo2<}
    \alr{\Delta \OO_{\rm do}^2}_{\rho_\theta^{(0)}} &= \alr{\OO_{\rm do}^2}_{\rho_\theta^{(0)}} = \sum_i b_i^2 + 2\sum_{1\le i<j\le N} b_i b_j \alr{\fY_{i+1,j}}_{\rho_\theta^{(0)}} \nonumber\\
    &\le \sum_i b_i^2 + 2\sum_{1\le i<j\le N} b_i b_j |a_i||a_j| \nonumber\\
    &\le \sum_i b_i^2 + \lr{\sum_i b_i |a_i|}^2.
\end{align}
Here $\fY_{i,j} := \bigotimes_{k=i}^j Y_k$, and its expectation value is only contributed by $\ket{1_i}$ and $\ket{1_j}$, which leads to the second line. Finally, choosing $b_i = |a_i|$, $\alr{\Delta \OO_{\rm do}^2}_{\rho_\theta^{(0)}}\le 1+1=2$ due to normalization, and \begin{equation}
    \partial_\theta\alr{\OO_{\rm do}}_{\rho_\theta^{(0)}} \ge c_{\rm do}vT,
\end{equation}
according to \eqref{suppeq:dphi>T} and our assumption \eqref{suppeq:a_i>c}. This leads to \eqref{suppeq:Odo>T2} in the noiseless case.

With noise, the (dual map of) dephasing channel acting on observable \eqref{suppeq:O_do} retains its form but with coefficients $b_i=|a_i|(1-2p)^i$. Then we can still use \eqref{suppeq:dphi>T} and \eqref{suppeq:Odo2<} with these coefficients, yielding \begin{equation}
    \partial_\theta\alr{\OO_{\rm do}}_{\rho_\theta} \ge c_{\rm do}vT (1-2p)^{2vT},
\end{equation}
and $\alr{\Delta \OO_{\rm do}^2}_{\rho_\theta}\le \alr{\Delta \OO_{\rm do}^2}_{\rho_\theta^{(0)}}\le 2$. This establishes \eqref{suppeq:Odo>T2} for noise $p<1/(vT)$ because $(1-2p)^{2vT}\ge (1-2p)^{2/p}\ge \ee^{-4}$ for $p<1/2$. 
\end{proof}

If $T\ll N$, \eqref{suppeq:a_i>c} can be verified by the exact solution $a_i\approx (-\ii)^{i-1}(i/t) J_i(vT)$ with $v=2$ \cite{domino_bessel} where $J_i$ is the Bessel function of the first kind. Here the exact solution comes from the fact that the restricted Hamiltonian in the 1-DW subspace is the solvable matrix \begin{equation}
    H^{(1)}_{ij}= \left\{\begin{array}{cc}
        1,  & |i-j|=1 \\
        0, & \text{otherwise}
    \end{array}\right. 
\end{equation}
In practice, \eqref{suppeq:a_i>c} holds all the way up to $vT\approx N$ from numerics, when the wavefront bounces off the right boundary.

Note that~\cite{metro_domino21} proposes another measurement scheme: after sensing, one further evolve the system under $H_{\rm do}$ for some time before the final on-site measurement, so that the DW bounces off the right boundary and approximately returns to the initial site $1$. However, this is not parametrically optimal, because the wave packet actually \emph{spreads} as Bessel function $a_i\propto (i/t) J_i(vT)= \order(T^{-1/3})\ll 1, \forall i$, yielding a small final signal $\alr{Y_1}_{\rm bounce}=\Theta( N^{-1/3})$ and $\partial_\theta \alr{Y_1}_{\rm bounce}=\Theta(N^{2/3})$ that leads to $\delta \theta =\Theta(N^{-2/3})$ 
Here $Y_1$ is the observable on the first site in the final state after bouncing off.

\subsection{Partitioned domino state}

For partitioned domino state $\ket{\Psi_{\rm do}}$ (\eqref{eq:ZiZj} in the main text), one wants to use \begin{equation}\label{suppeq:T=Lv} 
    T\approx L/(2v),
\end{equation}
with the DW velocity $v= 2$ so that DWs of each group propagate to the edges. To simplify the situation, we choose $T$ to be slightly smaller than \eqref{suppeq:T=Lv}, to make sure the leakage of each DW to other groups is small, so that the groups evolve almost independently. For example, for Fig.~3 in the main text, we choose \begin{equation}\label{suppeq:T=num}
    T=1.3,\quad 3.7,\quad 8.7,\quad  23.9,\quad 49,
\end{equation}
for the five values of $L=10,20,40,100,200$. In the numerical simulation, we simulate individual states $\ket{\Psi_{\rm simu}}$ generated from  single $|+\rangle$'s and approximate $\ket{\Psi_{\rm do}}$ using tensor product of these states. 
We estimate that such approximation of independent dynamics causes small error $\norm{|\Psi_{\rm do}\rangle-|\Psi_{\rm simu}\rangle^{\otimes N/L}}^2\lessapprox 2\%$ 
for the choices \eqref{suppeq:T=num}. 

Based on the $T$ chosen above, $\ket{\Psi_{\rm do}}$ is approximately $N/L$ copies of the $L$-qubit state \begin{equation}\label{suppeq:aij}
     \frac{1}{\sqrt{2}}\ket{\bm{0}}+\frac{1}{\sqrt{2}}\sum_{-L/2+1\le i,j\le L/2}a_{i,j} \ket{1_{i,j}},
\end{equation}
which is evolved from a single $\ket{+}$ at site $1$. 
Here $\ket{1_{i,j}}$ is all-$1$ from $i$ to $j$ (including these two sites), and all-$0$ elsewhere.
Since the Hamiltonian $H_{\rm do}$ is real, $a_{i,j}$ is either real for $|i-j|$ being even, or purely imaginary for odd $|i-j|$. This structure simplifies the observable for sensing $\theta\approx 0$: one just measures in the $y$ on-site basis, because for GHZ metrology $\ket{\widetilde{\GHZ}}=\ee^{-\ii\theta \sum_{i=1}^L Z_i} \frac{1}{\sqrt{2}}\lr{\ket{0^{\otimes L}} \pm \ii^{L+1} \ket{1^{\otimes L}} }$ at $\theta=0$, 
\begin{equation}
    \partial_\theta \alr{\fY_{1,L}}_{\widetilde{\GHZ}} = \pm L \lr{\ii \bra{\bm{0}}\fY_{1,L} \ii^{L+1} \ket{\bm{1}} -\ii \bra{\bm{1}}(-\ii)^{L+1} \fY_{1,L} \ket{\bm{0}}} = \mp 2L,
\end{equation}
is an optimal observable for any $L$. Note that the following still applies to cases beyond $\theta\approx 0$, where one may want to optimize numerically over the local phases $\phi_i$ of local observables $Y_i \rightarrow \cos(\phi_i)X_i+\sin(\phi_i)Y_i$. 

For sensing $\theta\approx 0$ with $N/L$ copies of \eqref{suppeq:aij}, we choose an observable \begin{equation}\label{suppeq:bY}
    \mathcal{O}_{\rm do}'=\sum_{-L/2+1\le i\le 1,1\le j\le L/2} b_{i,j}\fY_{i,j},
\end{equation}
for each $L$-qubit group with $b_{i,j}= |a_{i,j}|$, and the total observable would be a sum of $\mathcal{O}_{\rm do}'$ over groups, just like Theorem \ref{suppthm1}. This is equivalent to \eqref{eq:H=XZZ} in the main text, where 
$b_{i,j}=0$ for $(i,j)$ that does not appear in the sum \eqref{suppeq:bY}. 
The idea of \eqref{suppeq:bY} is similar to Theorem 3 in the main text, where we want the signal to be contributed from terms like $\bra{\bm{0}} \fY_{i,j}\ket{1_{i,j}}$. Note that one can also allow for terms with $i>1$ in \eqref{suppeq:bY}; our choice simplifies the calculation a bit. Still, it is tedious to perform a fully analytical treatment like Theorem 3 in the main text, so we numerically calculate the signal and variance of \eqref{suppeq:bY}, as shown in Fig.~3 in the main text. The results are as expected. 

\section{Generalizations}\label{sec:general}

\subsection{General QFI upper bound}\label{sec:generalnoise}


Here we derive a general QFI upper bound of $\Theta(N/p)$ for any multi-probe systems where each probe is independently and identically subject to a quantum signal and a quantum noise with noise probability $p$. Specifically, we consider the following two situations: 

\begin{enumerate}[(1)]
    \item The noise quantum channel $\mN_\theta$\footnote{Note that we use notations $\mN_\theta,\mN_\theta^{(0)}$ instead of $\mN_{\theta,p},\mN_{\theta,0}$ in the main text, because for a general channel there is no unique way to define the error probability $p$. We will define $p$ by a convenient choice shortly. } acting on each probe describes the evolution of an open quantum system with a Hamiltonian $H_{\rm p}$
    \begin{equation}
    \label{suppeq:noise-1}
        \mN_\theta = \exp(\theta \sH t + \gamma \sL t),
    \end{equation}
    where $t$ is the evolution time \footnote{Note that $t$ is different than the time $T$ ($T'$) to prepare locally generated resource states in Section \ref{sec:thm2}. Nevertheless, in order to achieve optimal noisy metrology, one needs to choose $T$ such that $\gamma t=\Theta(T^{-d})$ according to Theorem \ref{suppthm:Orev>}. Below we set $t = 1$ without loss of generality by rescaling the noise rate $\gamma$.}, $\sH(\rho) = -\ii [ H_{\rm p},\rho]$ describes the Hamiltonian evolution, and the noise is described by 
    \begin{equation}
        \sL(\rho) = \sum_{i=1}^k L_i \rho L_i^\dagger - \frac{1}{2} \{L_i^\dagger L_i,\rho\}.  
    \end{equation}
    $\theta$ and $\gamma$ represent the strengths of the signal and the noise, respectively, where $\theta$ is the unknown parameter to be detected. We consider the regime where $\gamma \ll 1$ and $t=1$. The noise probability $p$ is defined to be 
    \begin{equation}
        p = 1 - \ee^{-\gamma t},
    \end{equation}  
    and we have $p = \Theta(\gamma)$. 
    Viewing $\theta/\gamma$ as the parameter to be estimated, we can use results from previous works~\cite{zhou2018achieving,metro_QEC17,wan2022bounds} to obtain 
    \begin{align}
    \label{suppeq:upper-1}
    \mF(\mN_\theta^{\otimes N}) &\leq \left(\frac{1}{\gamma^2}\right) \cdot 4 \min_{\substack{\forall h,\vh,\frakh, \; s.t.,\\H + h_{00}\id + \vh^\dagger \vL + \vh \vL^\dagger + \vL^\dagger \frakh \vL = 0}}\norm{(\vh \id + \frakh \vL)^\dagger(\vh \id + \frakh \vL)} \cdot (N\gamma t) \nonumber\\ &= \order\left(\frac{N}{\gamma}\right) = \order\left(\frac{N}{p}\right), 
    \end{align}
    where $\vL = \begin{pmatrix}
        L_1 \\ \vdots \\ L_r
    \end{pmatrix}$, $h_{00} \in \mathbb{R}$, $\vh \in \mathbb{C}^r$, $\frakh \in \mathbb{C}^{r \times r}$ and is Hermitian. Here $\mF(\mN_\theta^{\otimes N})$ is the channel QFI defined by $\mF(\mN_\theta^{\otimes N})(\rho)$ maximized over all input states $\rho$. 
    The only assumption is the channel violates the so-called ``Hamiltonian-not-in-Lindblad-span'' condition, i.e., $H \in {\rm span}\{\id,L_i,L_i^\dagger,L_i^\dagger L_j,\forall i,j\}$. The condition is violated for, e.g., Pauli-Z signal and dephasing noise, as discussed in the main text. This situation also encompasses many commonly encountered noise types, e.g., depolarizing noise, amplitude damping noise, erasure noise, etc. When the condition holds, the HL is in principle achievable using techniques like quantum error correction~\cite{zhou2018achieving}, which is beyond the scope of our work. 
    \item Another broadly discussed scenario is the following, where the evolution of each probe is described by 
    \begin{equation}
    \label{suppeq:noise-2}
    \mN_\theta = \exp(\gamma \sL t) \exp(\theta \sH t),
    \end{equation}
    where $\sL$ and $\sH$ are defined as above, which is the composition of a Hamiltonian evolution and a noise channel afterwards. Note that situation (2) is the same as situation (1) when $\sH$ and $\sL$ commutes, e.g., Pauli-Z signal with depolarizing noise. The assumption that the noise acts after the Hamiltonian evolution is less commonly encountered in practice, but can be much simpler to deal with in theory when the noise channel $\exp(\gamma \sL t)$ is analytically solvable, see, e.g.~\cite{metro_noise12}. 
    Again, we consider the limiting case where $\gamma \ll 1$ and $t$ is a constant and $p$ is the noise probability that satisfies $p = \Theta(\gamma)$. Note that here the noise channel $\mN = \exp(\gamma \sL t)$ can represent any CPTP map, i.e., any noisy quantum channel, and we further require the ``Hamiltonian-not-in-Kraus-span'' condition to be violated, i.e, $H \in {\rm span}\{\id,K_i^\dagger K_j,\forall i,j\}$. The assumption again holds in most noisy quantum systems with e.g., depolarizing noise, amplitude damping noise, etc. We assume $\mN(\cdot) = \sum_{i=0}^{r} K_i (\cdot) K_i^\dagger$, where $K_i$ are Kraus operators that we specify later. From previous works, the QFI is upper bounded by~\cite{metro_noise12,channel_estimate21}
    \begin{equation}
    \mF(\mN_\theta^{\otimes N}) \leq 4 \min_{\forall h, \; s.t., Ht + \vK^\dagger h \vK = 0} \norm{\vK^\dagger h^2 \vK - H^2t^2} N,
    \end{equation}
    where $\vK = \begin{pmatrix}
        K_0  \\ \vdots \\ K_{r}
    \end{pmatrix}$, $h \in \mathbb{C}^{(r+1)\times(r+1)}$ is Hermitian. Expanding $\vK$ and $\gamma h$ in Taylor expansions of $\gamma t$, we have 
    \begin{gather}
        K_0 = \id - \frac{{\gamma t}}{2}\sum_{i=1}^r L_i^\dagger L_i + \order((\gamma t)^{3/2}), \\
        K_i = \sqrt{\gamma t} L_i + \order(\gamma t), \quad i=1,\ldots,r,\\
        \gamma h =  \begin{pmatrix}
         h_{00}{\gamma t}+ \order((\gamma t)^{3/2})   &  \vh^\dagger (\gamma t)^{1/2} + \order(\gamma t)\\
         \vh (\gamma t)^{1/2} + \order(\gamma t)   & \frakh + \order((\gamma t)^{1/2}) 
         \label{suppeq:cond-1}\\
        \end{pmatrix},
    \end{gather}
    such that $Ht + \vK^\dagger h \vK = 0$ translates to 
    \begin{equation}
        H + h_{00} \id + \vh^\dagger \vL + \vL^\dagger \vh + \frakh \vL^\dagger \vL = 0, \label{suppeq:cond-2}
    \end{equation}
    where $h_{00} \in \mathbb{R}$, $\vh \in \mathbb{C}^{r}$, $\frakh \in \mathbb{C}^{r\times r}$ and is Hermitian. Then have  
    \begin{multline}
    \min_{\forall h, \; s.t., Ht + \vK^\dagger h \vK = 0} \norm{\vK^\dagger h^2 \vK - H^2t^2} \leq \\ \frac{1}{\gamma^2} \bigg( (\gamma t) \min_{\substack{\forall h,\vh,\frakh, \; s.t.,\\H + h_{00}\id + \vh^\dagger \vL + \vh \vL^\dagger + \vL^\dagger \frakh \vL = 0}}\norm{(\vh \id + \frakh \vL)^\dagger(\vh \id + \frakh \vL)}  + \order((\gamma t)^{3/2}) \bigg)
    \end{multline}
    where we ignored the $H^2t^2$ term because $\norm{A-B} \leq \norm{A}$ when $A-B$ and $B$ are arbitrary positive semidefinite matrices. It implies 
    \begin{equation}
        \mF(\mN_\theta^{\otimes N}) \leq \frac{N t}{\gamma} \cdot \text{constant} + \order\left(\frac{N}{\gamma^{1/2}}\right) = \order\left(\frac{N}{p}\right),
    \end{equation}
    where the constant above is the same as the one appearing in \eqref{suppeq:upper-1}. 
\end{enumerate}

As discussed above, for general noise, including \eqref{suppeq:noise-1} and \eqref{suppeq:noise-2}, we always have a QFI upper bound of $\order(N/p)$. We show next how to generalize our theorems to find metrologically optimal states achieving this scaling. We will use the following two properties 
\begin{equation}
\label{suppeq:property}
    \norm{\partial_\theta \mN_\theta - \partial_\theta \mN^{(0)}_\theta}_\diamond = \order(p),\quad \norm{\mN_\theta - \mN^{(0)}_\theta}_\diamond = \order(p),
\end{equation}
where $^{(0)}$ represents the noiseless channel when taking $\gamma = 0$ (i.e., $\mN^{(0)}_\theta = \exp(\theta \sH t)$), $\norm{\cdot}_\diamond$ is the diamond norm, and \eqref{suppeq:property} is satisfied for both \eqref{suppeq:noise-1} and \eqref{suppeq:noise-2}. Here the correctness of \eqref{suppeq:property} can be seen from the two expressions of the noisy channels (\eqref{suppeq:noise-1} and \eqref{suppeq:noise-2}), $\mN^{(0)}_\theta = \mN_\theta \big|_{\gamma = 0}$ and the fact that $\gamma = \Theta (p)$. 

One example of the above formalism of noise channel is the amplitude damping noise (by default we set the evolution time $t = 1$). We assume 
\begin{gather}
    \mN_\theta = \exp(\gamma \sL) \exp(\theta \sH),\\
    \sL(\cdot) = \gamma ( \ket{0}\bra{1}(\cdot)\ket{1}\bra{0} - \frac{1}{2}\{\ket{1}\bra{1},(\cdot)\}). 
    \end{gather}
    Or equivalently, 
    \begin{gather}
    \mN_\theta(\cdot) = K_1 e^{-iH_{\rm p}\theta}(\cdot) e^{iH_{\rm p}\theta} K_1^\dagger + K_2 e^{-iH_{\rm p}\theta}(\cdot) e^{iH_{\rm p}\theta} K_2^\dagger,\\
    K_1 = \ket{0}\bra{0} + \sqrt{1 - p} \ket{1}\bra{1},\quad 
    K_2 = \sqrt{p}\ket{0}\bra{1}. 
\end{gather}
where $p = 1 - e^{-\gamma}$. In the noise regime we consider, i.e., $1/N < p < 1/2$, we have $\gamma/2 \leq p \leq \gamma$ and clearly $p = \Theta(\gamma)$ as required.

\subsection{General optimality of states constructed in the main text}

The mapping from any qudit system to a qubit system is straightforward.
To estimate $\theta$ in \eqref{suppeq:noise-1} or \eqref{suppeq:noise-2}, we can consider the two-dimensional subspace spanned by two eigenstates corresponding to the highest and the lowest eigenvalues of $H$ and denote them by $\ket{0}$ and $\ket{1}$. We further assume, without loss of generality, the highest and lowest eigenvalues are $1$ and $-1$ and $t = 1$. Then any multi-qubit state described in the main text for sensing under dephasing noise naturally represents a multi-probe state in the general context here (\eqref{suppeq:noise-1} or \eqref{suppeq:noise-2}). We show the state optimality below. 

\begin{thm}\label{suppthm:generalize}
    Theorem~\ref{suppthm1},\ref{suppthm:Orev>},\ref{suppthm:localM} continue to hold for the generalized situations above. 
\end{thm}

\begin{proof}
We first deal with Theorem~\ref{suppthm:Orev>} that is conceptually simpler. In its proof, the only step related to the noise channel is the final paragraph. Since the variance bound still holds for general single-qubit noise that cannot create long-range correlation, we only need to modify the derivative bound \eqref{suppeq:P-P0<}. In a nutshell, \eqref{suppeq:P-P0<} generalizes because it only uses the fact that $\mN_\theta$ is $p$-close to $\mN_\theta^{(0)}$ in the sense of \eqref{suppeq:property}, not the specific form of $\mN$.

More precisely, below we use $\mN_\theta^\dagger,\mN_\theta^{(0)\dagger}$ to represent the dual maps of $\mN_\theta,\mN_\theta^{(0)}$. Then the $p$-closeness \eqref{suppeq:property} for the original channels implies 
\begin{equation}
\label{suppeq:property_dual}
    \norm{(\partial_\theta\mN^\dagger_\theta\otimes {\id})(\OO) - (\partial_\theta\mN^{(0)\dagger}_\theta\otimes {\id})(\OO)} \le \order(p)\norm{\OO},\quad \norm{(\mN_\theta^\dagger\otimes {\id})(\OO) - (\mN^{(0)\dagger}_\theta\otimes {\id})(\OO)} \le \order(p)\norm{\OO},
\end{equation}
where ${\id}$ is the identity channel acting on ancillary degrees of freedom, and $\OO$ is any operator acting on the whole system.

We can modify \eqref{suppeq:P-P0<} as follows (we ignore the tensor product with $\id$ operators below for simplicity),
\begin{align}\label{suppeq:P-P0<1}
    &\abs{ \partial_\theta \lr{\alr{\widetilde{P}_I}_{\rho_\theta}- \alr{\widetilde{P}_I}_{\rho_\theta^{(0)}}} }= \abs{ \partial_\theta \alr{\mN_\theta^{\dagger \otimes N}(\widetilde{P}_I)-\mN_\theta^{\dagger  (0)\,\otimes N}(\widetilde{P}_I) }_{\rho} } \nonumber\\
    &= \abs{\sum_i \alr{(\partial_\theta \mN^\dagger_\theta)_i \otimes \mN_\theta^{\dagger\otimes N-1}(\widetilde{P}_I)-(\partial_\theta \mN_\theta^{(0)\dagger})_i \otimes\mN_\theta^{(0)\dagger\,\otimes N-1}(\widetilde{P}_I) }_{\rho} } \nonumber\\
    &\le \sum_{i\in \widetilde{I}_T} \abs{ \alr{(\partial_\theta \mN^\dagger_\theta)_i \otimes \mN_\theta^{\dagger\otimes M'-1}(\widetilde{P}_I)-(\partial_\theta \mN_\theta^{(0)\dagger})_i \otimes\mN_\theta^{(0)\dagger\,\otimes M'-1}(\widetilde{P}_I) }_{\rho} } + \order(L^2\epsilon) \nonumber\\
    &\le M'\max_i\norm{(\partial_\theta \mN_\theta^\dagger)_i \otimes \mN_\theta^{\dagger\otimes M'-1}(\widetilde{P}_I)-(\partial_\theta \mN_\theta^{(0)\dagger})_i \otimes\mN_\theta^{\dagger\otimes M'-1}(\widetilde{P}_I) } \nonumber\\
    &\quad + M'\max_i\sum_{g=1}^{M'-1}  \norm{(\partial_\theta \mN_\theta^{(0)\dagger})_i \otimes \mlr{ \mN_\theta^{\dagger\,\otimes g}\otimes \mN_\theta^{(0)\dagger\,\otimes M'-1-g}(\widetilde{P}_I)-\mN_\theta^{\dagger\,\otimes g-1}\otimes \mN_\theta^{(0)\dagger\,\otimes M'-g}(\widetilde{P}_I)} }+\order(L^2\epsilon) \nonumber\\
    &\le M'\cdot M'\order(p)+\order(L^2\epsilon)\le \Theta(L^4) p+\order(L^2\epsilon).
\end{align} 
Here in the third line, we have again used that $\widetilde{P}_I$ is almost supported in region $\widetilde{I}_T$ with size $M'=\Theta(L^2)$: For a faraway site $i$, $(\partial_\theta\mN_\theta)_i$ acts on identity operator and becomes zero. To get the last line, we have used \eqref{suppeq:property_dual} with the triangle inequality. The generalized Theorem~\ref{suppthm:Orev>} is then proven.

Theorem~\ref{suppthm:localM} generalizes in a similar way.

It remains to generalize Theorem~\ref{suppthm1}. In its proof, \eqref{suppeq:FI>G2} still holds because \eqref{suppeq:FI-FI0} only uses the continuity of QFI and the $p$-closeness \eqref{suppeq:property} for $\mN_\theta=\DD_{\theta,p}$ and $\mN_\theta^{(0)}=\DD_{\theta,0}$. Then we still use Lemma~\ref{lem:normO} to choose an observable $\OO_I$ for each group. Choose $\sigma_\theta = \rho_{I,\theta}$ and $A = H_I := \sum_{i\in I} H_i$ in Lemma~\ref{lem:normO}, where $H_i$ represents the signal Hamiltonian on each probe, it was shown in the proof of Theorem~\ref{suppthm1} that when \eqref{suppeq:B<pM} holds (which we will prove below), there exists $\OO_I$ that satisfies \eqref{suppeq:oi-1} and \eqref{suppeq:oi-2}, and the rest of the proof also applies to general signal and noise. 

To prove \eqref{suppeq:B<pM} for general signal and noise, 
we note by direct calculation,
\begin{align}
    B&=\partial_\theta\rho_{I,\theta} + \ii [H_I, \rho_{I,\theta}] = \partial_\theta\lr{\mN_\theta^{\otimes \G} (\rho_I)} + \ii [H_I, \mN_\theta^{\otimes \G} (\rho_I) ]\nonumber\\
    &= \sum_{i\in I} \mN_\theta^{\otimes \G-1} \otimes \lr{\partial_\theta \mN_\theta}_i (\rho_I) + \ii [H_i, \mN_\theta^{\otimes \G}(\rho_I)] = \sum_{i\in I} \mN_\theta^{\otimes \G-1}\lr{ \lr{\partial_\theta \mN_\theta}_i(\rho_I) + \ii [H_i, \lr{\mN_\theta}_i(\rho_I)]},
\end{align}
where we have used the Leibniz product rule to get the second line, and that $H_i$ commutes with channels acting on sites other than $i$ in the end. Taking trace norm, we have \begin{align}
    &\norm{B}_1\le \G \max_i \norm{\lr{\partial_\theta \mN_\theta}_i (\rho_I) + \ii [H_i, \lr{\mN_\theta}_i(\rho_I)]}_1 \nonumber\\
    &= \G \max_i \norm{\lr{\partial_\theta \mN_\theta - \partial_\theta \mN_\theta^{(0)}}_i(\rho_I) + \lr{\partial_\theta \mN_\theta^{(0)}}_i (\rho_I) + \ii \Big[H_i, \lr{\mN_\theta^{(0)}}_i (\rho_I)\Big]
    +\ii \Big[H_i, \lr{\mN_\theta-\mN_\theta^{(0)}}_i (\rho_I)\Big]}_1 \nonumber\\
    &= \G \max_i \norm{\lr{\partial_\theta \mN_\theta - \partial_\theta \mN_\theta^{(0)}}_i(\rho_I) + \ii \Big[H_i, \lr{\mN_\theta-\mN_\theta^{(0)}}_i(\rho_I)\Big] }_1 =\G\cdot \order(p).
\end{align}
Here the middle two terms in the second line cancels because $\lr{\partial_\theta \mN_\theta^{(0)}}_i \rho_I =- \ii \Big[H_i, \lr{\mN_\theta^{(0)}}_i \rho_I\Big]$, and we have used \eqref{suppeq:property} together with $\norm{H_i}= \order(1)$ in the end. This proves \eqref{suppeq:B<pM} and generalizes Theorem~\ref{suppthm1}.

\end{proof}

We expect our results generalize even beyond independent signal and noise on each probe to include signal and noise that might be correlated within each group of probes, because, for example, the generalized Theorem~\ref{suppthm1} only relies on the continuity of QFI and the existence of an observable in each group saturating its QFI. 

In addition, our protocol may also tolerate noise in state preparation and measurement when the noise probability per group $p_{\text{per group}}$ is smaller than some constant $c_{\text{per group}}$: Take Theorem~\ref{suppthm:Orev>} with measurement noise for example. Suppose the measurement noise is characterized by a channel $\mathcal{E}$ prior to the ideal measurement. The ideal measurement observable $\widetilde{P}_I$ for each group becomes $\mathcal{E}^\dagger\lr{\widetilde{P}_I}$. 
Let $\eta$ represents the error bound \eqref{suppeq:P-P0<1} for replacing $\rho_\theta\rightarrow \rho_\theta^{(0)}$, the additional error caused by $\mathcal{E}$ is 
\begin{align}\label{suppeq:measure_error}
    \abs{\partial_\theta\alr{\mathcal{E}^\dagger\lr{\widetilde{P}_I}-\widetilde{P}_I}_{\rho_\theta } } &\le \abs{\partial_\theta\alr{\mathcal{E}^\dagger\lr{\widetilde{P}_I}-\widetilde{P}_I}_{\rho_\theta^{(0)} } } + 2 \eta = \abs{\alr{\mlr{H,\mathcal{E}^\dagger\lr{\widetilde{P}_I}-\widetilde{P}_I}}_{\rho_\theta^{(0)}  } } + 2 \eta \nonumber\\
    &\le 2\G' \norm{\mathcal{E}^\dagger\lr{\widetilde{P}_I}-\widetilde{P}_I} + 3\eta \le 2\G'\times p_{\text{per group}}+3\eta, 
\end{align}
where $p_{\text{per group}}$ is the diamond-norm distance between $\mathcal{E}^\dagger$ and the identity channel. Here we changed $2\eta\rightarrow 3\eta$ to account for the error restricting $H:=\sum_i H_i$ to $H_{\widetilde{I}_T}$. \eqref{suppeq:measure_error} is then bounded by a constant times $\G'$, which is exactly the same as \eqref{suppeq:P-P0<1}, so the presence of $\mathcal{E}$ does not change the scaling of the final precision (because obviously the variance bound \eqref{suppeq:DeltaO<} is also robust). Note that for this generalization to hold, we need the observable to have a small operator norm $\norm{\widetilde{P}_I}=1$ comparing to $\partial_\theta \alr{\widetilde{P}_I}=\Theta(\G)$; that does not hold in the case of Theorem~\ref{suppthm1} (see \eqref{suppeq:dsigma=A} and \eqref{suppeq:L<A}). 

\section{Spin-squeezed states as metrological optimal states}

In this work we have mainly focused on optimal states given by Theorem 1 that have large intra-group and small inter-group correlations. However, there can be other classes of optimal states with different correlation structure. \cite{channel_estimate21,ulam2001spin,brask2015improved} studied a particular spin-squeezed state (SSS) sensing the channel \eqref{eq:Domega} in the main text, and found optimal performance even if every pair of qubits shares $\Theta(1)$ correlation. Below we prove that very general SSS are optimal for sensing magnetic field under any general single-qubit noise. 

\begin{thm}[Restatement of Theorem 4 in the main text]
\label{suppthm:squeeze}
    Suppose $\rho$ is spin-squeezed in the sense that \begin{align}\label{suppeq:squeeze1}
    \alr{X}_\rho &= \Theta(N), \\
        \alr{\Delta Y^2}_\rho &= N \xi^2, \label{suppeq:squeeze2} \\
        \alr{\Delta S^2}_\rho &= \order(N/\xi^2),\quad \forall S:=\sum_i S_i \text{ with } \norm{S_i}=1 \label{suppeq:squeeze3}
    \end{align}
    where $X=\sum_i X_i$ etc. 
    Each qubit undergoes a channel $\mN_\theta$ satisfying \eqref{suppeq:property} with $\mN_\theta^{(0)}(\cdot)=\ee^{-\ii\theta Z}(\cdot)\ee^{\ii\theta Z}$ for some $\xi < 1$. Then if $p=\order(\xi^2)$, the final state $\rho_\theta$ has the parametrically the same metrological power as its noiseless limit at $\theta = 0$, because \begin{equation}\label{suppeq:noisy_sq}
        \mathcal{F}(\rho_\theta) \geq \frac{1}{\alr{\Delta Y^2}_{\rho_\theta}}\lr{\partial_\theta\alr{Y}_{\rho_\theta}}^2\Big|_{\theta=0} = \Theta( N/\xi^2).
    \end{equation}
\end{thm}

\eqref{suppeq:noisy_sq} then saturates the optimal bound $\Theta(N/p)$ at $p=\Theta(\xi^2)$. Furthermore, $Y$ is an optimal observable here, indicating the existence of an optimal on-site measurement strategy. Note that here we must have $\xi = \Omega(1/\sqrt{N})$, because otherwise \eqref{suppeq:squeeze1} and \eqref{suppeq:squeeze2} cannot be simultaneously satisfied due to the uncertainty relation. 
Here we focus on sensing at $\theta = 0$ without loss of generality, because other values $\theta$ can be dealt with similarly by rotating the final observable $Y\rightarrow Y \cos(2\theta)+X\sin(2\theta)$. The conditions
\eqref{suppeq:squeeze1} and \eqref{suppeq:squeeze2} are standard for SSS, which simply mean large polarization along $x$ and small fluctuation in polarization along $y$.
Since $\alr{\Delta Y^2}=N$ for a product state along $x$-direction, the $\xi^2\ll 1$ is roughly the squeezing parameter. Taking $S=Z$ for example, the third condition \eqref{suppeq:squeeze3} requires that fluctuation in other directions cannot be much larger than what is predicted from \eqref{suppeq:squeeze2} and the saturation of the uncertainty relation $\alr{\Delta Y^2} \alr{\Delta Z^2} \geq \abs{\alr{X}}^2 =  \Omega (N^2)$. 

\begin{proof}
Acting the dual maps $\mN^\dagger_\theta$ on the observable, we have \begin{align}\label{suppeq:Y=}
&\alr{Y}_{\rho_\theta} = \sum_i \alr{\mN^\dagger_\theta (Y_i)}_\rho=\sum_i \alr{\mN_\theta^{(0)\dagger}(Y_i)}_\rho + \alr{(\mN^\dagger_\theta-\mN_\theta^{(0)\dagger})(Y_i)}_\rho = \sum_i \alr{\ee^{\ii\theta Z_i}Y_i \ee^{-\ii\theta Z_i}}_\rho + \order(pN), \\
&\partial_\theta\alr{Y}_{\rho_\theta}|_{\theta=0} = 2\sum_i \alr{X_i}_\rho + \order(pN)=\Theta(N), \label{suppeq:dY=}
\end{align}
where we have used \eqref{suppeq:property_dual}, and plugged in \eqref{suppeq:squeeze1} in the second line. Similarly, 
\begin{align}\label{suppeq:Y2=}
&\alr{Y^2}_{\rho_\theta}|_{\theta=0} = N+\sum_{i\neq j} \alr{\mN_0^\dagger(Y_i)\otimes \mN_0^\dagger(Y_j)}_\rho = N+ \alr{\lr{\sum_i\mN_0^\dagger(Y_i)}^2}_\rho -\sum_i \alr{\lr{\mN_0^\dagger(Y_i)}^2}_\rho \nonumber\\
&= \alr{\mlr{\sum_i \lr{Y_i+\order(p) S_i }}^2}_\rho + \sum_i \alr{\id-\lr{Y_i+\order(p) S_i}^2}_\rho = \alr{\mlr{\sum_i \lr{Y_i+\order(p) S_i }}^2}_\rho + \order(pN), 
\end{align}
where we have introduced an operator $S_i\propto \mN_0^\dagger(Y_i)-Y_i$ on spin $i$ with $\norm{S_i}=1$, and used $\id-Y_i^2=0$. 
Combining \eqref{suppeq:Y=} (the second term below) and \eqref{suppeq:Y2=},
\begin{align}
\alr{\Delta Y^2}_{\rho_\theta}|_{\theta=0} &= \alr{\mlr{\sum_i \lr{Y_i+\order(p)S_i}}^2}_\rho - \mlr{\alr{\sum_i \lr{Y_i+\order(p)S_i}}_\rho}^2 +\order(pN) \nonumber\\
&= \alr{\Delta Y^2}_\rho + \order(p) \mlr{C_\rho(Y,S)+C_\rho(S,Y)}+ \order(p^2) \alr{\Delta S^2}_\rho + \order(pN)\nonumber\\
&= N\xi^2 + \order(\, p \sqrt{\alr{\Delta Y^2} \alr{\Delta S^2} }\,)  + \order(p^2N/\xi^2) + \order(pN) = N\xi^2 +\order(N(p+p^2/\xi^2)) \nonumber\\
&=\Theta(N\xi^2), \quad \mathrm{if}\quad p=\order(\xi^2),
\label{suppeq:DeltaY2=}
\end{align}
where we have used Cauchy-Schwarz inequality and \eqref{suppeq:squeeze3} in the third line. Combining \eqref{suppeq:dY=} with \eqref{suppeq:DeltaY2=} yields \eqref{suppeq:noisy_sq}.

\end{proof}

One example of optimal SSS's that saturate the conditions in Theorem~\ref{suppthm:squeeze} are SSS's generated by one-axis twisting~\cite{channel_estimate21,ulam2001spin,squeez_93,brask2015improved}, 
\begin{equation}
\label{suppeq:sss-example}
    \ket{\psi_{\mu,\nu}} = \exp(-i \nu X/2) \exp(-i \mu Z^2/8 ) \exp(-i\pi Y/4) \ket{\bm{0}},  
\end{equation}
and taking $\xi = \sqrt{p}$ and
\begin{equation}
    \mu = \frac{1}{N\xi},\quad \nu = \frac{\pi}{2} - \frac{1}{2} \arctan \frac{4\sin(\mu/2)\cos^{N-2}(\mu/2)}{1 - \cos^{N-2}\mu}. 
\end{equation}
In this case, let $a = 1 - \cos^{N-2} \mu$, $b = 4 \sin(\mu/2)\cos^{N-2}(\mu/2)$, $\rho = \ket{\psi_{\mu,\nu}}\bra{\psi_{\mu,\nu}}$, we have  
\begin{gather}
    \braket{X}_\rho = N\cos\left(\frac{\mu}{2}\right)^{N-1} = \Theta(N), \;\;\braket{Y}_\rho = 0,\;\; \braket{Z}_\rho = 0, \\
    \braket{\Delta X^2}_\rho = N \left( N (1 - \cos^{2(N-1)}\left(\frac{\mu}{2}\right) - (N-1)\frac{a}{2}\right) = \order(1/\xi^4),\\
    \braket{\Delta Y^2}_\rho = N \left(1 + \frac{1}{4}(N-1)(a - \sqrt{a^2+b^2})\right) = \Theta(N \xi^2),\\
    \braket{\Delta Z^2}_\rho = N \left(1 + \frac{1}{4}(N-1)(a + \sqrt{a^2+b^2})\right) = \Theta(N/\xi^2).\label{suppeq:SSS-example-Z-variance}
\end{gather}
\eqref{suppeq:squeeze1} and \eqref{suppeq:squeeze2} hold true from above, and \eqref{suppeq:squeeze3} further holds because let $S = s_x X + s_y Y + s_z Z$, we have $\braket{\Delta S^2}_\rho \leq s_x^2 \braket{\Delta X^2}_\rho + s_y^2 \braket{\Delta Y^2}_\rho + s_z^2 \braket{\Delta Z^2}_\rho + 2\abs{s_xs_y}\sqrt{\braket{X^2}_\rho\braket{Y^2}_\rho} + 2\abs{s_ys_z}\sqrt{\braket{Y^2}_\rho\braket{Z^2}_\rho} + 2\abs{s_zs_x}\sqrt{\braket{Z^2}_\rho\braket{X^2}_\rho} \leq \order(N/\xi^2)$.

Note that the spin-squeezed states have a different correlation structure from the short-ranged correlated states we introduced previously. Take \eqref{suppeq:sss-example} as an example. It is a permutation invariant state. We consider a sufficiently small $\xi = \sqrt{p}$ such that $\braket{\Delta Z^2}_\rho / N$ is bounded below by a constant larger than 1. 
Then \eqref{suppeq:SSS-example-Z-variance} implies $\braket{Z_iZ_j}_\rho - \braket{Z_i}_\rho\braket{Z_j}_\rho = 1/ \Theta(N\xi^2)$ when $i \neq j$. Thus, $C_\rho(I;J)$ in \eqref{suppeq:sumJCIJ<'} at least $1/\Theta(N\xi^2)$ when $I \neq J$, and $\sum_{J: J\notin \mathcal{S}_I} C_\rho(I;J) = \Omega(1/M\xi^2)=\Omega(M^{-1})$ because only a finite number of $J$s in $\mathcal{S}_I$ are excluded in the sum. 
Therefore, the spin-squeezed state does not satisfy \eqref{suppeq:sss-example}. 


The above proofs can generalize to cover signal/noise that acts independently but not identically on each qubit. Finally, we note that for purely dephasing channel (\eqref{eq:Domega} in the main text), we can take $S_i=Y_i$ in the proof, and one does not require condition \eqref{suppeq:squeeze3} any more.

\end{appendix}

\bibliographystyle{quantum}
\bibliography{biblio}

\providecommand{\noopsort}[1]{}\providecommand{\singleletter}[1]{#1}%
\begin{thebibliography}{10}

\bibitem{metro_rev11}
Vittorio Giovannetti, Seth Lloyd, and Lorenzo Maccone.
\newblock ``Advances in quantum metrology''.
\newblock \href{https://dx.doi.org/10.1038/nphoton.2011.35}{Nature photonics {\bf 5}, 222--229}~(2011).

\bibitem{sensing_rmp}
C.~L. Degen, F.~Reinhard, and P.~Cappellaro.
\newblock ``Quantum sensing''.
\newblock \href{https://dx.doi.org/10.1103/RevModPhys.89.035002}{Rev. Mod. Phys. {\bf 89}, 035002}~(2017).

\bibitem{pezze2018quantum}
Luca Pezz\`e, Augusto Smerzi, Markus~K. Oberthaler, Roman Schmied, and Philipp Treutlein.
\newblock ``Quantum metrology with nonclassical states of atomic ensembles''.
\newblock \href{https://dx.doi.org/10.1103/RevModPhys.90.035005}{Rev. Mod. Phys. {\bf 90}, 035005}~(2018).

\bibitem{pirandola2018advances}
Stefano Pirandola, Bhaskar~Roy Bardhan, Tobias Gehring, Christian Weedbrook, and Seth Lloyd.
\newblock ``Advances in photonic quantum sensing''.
\newblock \href{https://dx.doi.org/10.1038/s41566-018-0301-6}{Nat. Photonics. {\bf 12}, 724}~(2018).

\bibitem{caves1981quantum}
Carlton~M. Caves.
\newblock ``Quantum-mechanical noise in an interferometer''.
\newblock \href{https://dx.doi.org/10.1103/PhysRevD.23.1693}{Phys. Rev. D {\bf 23}, 1693--1708}~(1981).

\bibitem{yurke19862}
Bernard Yurke, Samuel~L. McCall, and John~R. Klauder.
\newblock ``Su(2) and su(1,1) interferometers''.
\newblock \href{https://dx.doi.org/10.1103/PhysRevA.33.4033}{Phys. Rev. A {\bf 33}, 4033--4054}~(1986).

\bibitem{ligo2011gravitational}
{LIGO Collaboration}.
\newblock ``A gravitational wave observatory operating beyond the quantum shot-noise limit''.
\newblock \href{https://dx.doi.org/10.1038/nphys2083}{Nature Physics {\bf 7}, 962--965}~(2011).

\bibitem{ligo2013enhanced}
{LIGO Collaboration}.
\newblock ``Enhanced sensitivity of the ligo gravitational wave detector by using squeezed states of light''.
\newblock \href{https://dx.doi.org/10.1038/nphoton.2013.177}{Nature Photonics {\bf 7}, 613--619}~(2013).

\bibitem{le2013optical}
David Le~Sage, Koji Arai, David~R Glenn, Stephen~J DeVience, Linh~M Pham, Lilah Rahn-Lee, Mikhail~D Lukin, Amir Yacoby, Arash Komeili, and Ronald~L Walsworth.
\newblock ``Optical magnetic imaging of living cells''.
\newblock \href{https://dx.doi.org/10.1038/nature12072}{Nature {\bf 496}, 486--489}~(2013).

\bibitem{lemos2014quantum}
Gabriela~Barreto Lemos, Victoria Borish, Garrett~D Cole, Sven Ramelow, Radek Lapkiewicz, and Anton Zeilinger.
\newblock ``Quantum imaging with undetected photons''.
\newblock \href{https://dx.doi.org/10.1038/nature13586}{Nature {\bf 512}, 409--412}~(2014).

\bibitem{tsang2016quantum}
Mankei Tsang, Ranjith Nair, and Xiao-Ming Lu.
\newblock ``Quantum theory of superresolution for two incoherent optical point sources''.
\newblock \href{https://dx.doi.org/10.1103/PhysRevX.6.031033}{Physical Review X {\bf 6}, 031033}~(2016).

\bibitem{abobeih2019atomic}
MH~Abobeih, J~Randall, CE~Bradley, HP~Bartling, MA~Bakker, MJ~Degen, M~Markham, DJ~Twitchen, and TH~Taminiau.
\newblock ``Atomic-scale imaging of a 27-nuclear-spin cluster using a quantum sensor''.
\newblock \href{https://dx.doi.org/10.1038/s41586-019-1834-7}{Nature {\bf 576}, 411--415}~(2019).

\bibitem{wineland1992spin}
D.~J. Wineland, J.~J. Bollinger, W.~M. Itano, F.~L. Moore, and D.~J. Heinzen.
\newblock ``Spin squeezing and reduced quantum noise in spectroscopy''.
\newblock \href{https://dx.doi.org/10.1103/PhysRevA.46.R6797}{Phys. Rev. A {\bf 46}, R6797--R6800}~(1992).

\bibitem{bollinger1996optimal}
J.~J. Bollinger, Wayne~M. Itano, D.~J. Wineland, and D.~J. Heinzen.
\newblock ``Optimal frequency measurements with maximally correlated states''.
\newblock \href{https://dx.doi.org/10.1103/PhysRevA.54.R4649}{Phys. Rev. A {\bf 54}, R4649--R4652}~(1996).

\bibitem{leibfried2004toward}
D.~Leibfried, M.~D. Barrett, T.~Schaetz, J.~Britton, J.~Chiaverini, W.~M. Itano, J.~D. Jost, C.~Langer, and D.~J. Wineland.
\newblock ``Toward heisenberg-limited spectroscopy with multiparticle entangled states''.
\newblock \href{https://dx.doi.org/10.1126/science.1097576}{Science {\bf 304}, 1476--1478}~(2004).

\bibitem{taylor2008high}
JM~Taylor, Paola Cappellaro, L~Childress, Liang Jiang, Dmitry Budker, PR~Hemmer, Amir Yacoby, R~Walsworth, and MD~Lukin.
\newblock ``High-sensitivity diamond magnetometer with nanoscale resolution''.
\newblock \href{https://dx.doi.org/10.1038/nphys1075}{Nature Physics {\bf 4}, 810--816}~(2008).

\bibitem{zhou2020quantum}
Hengyun Zhou, Joonhee Choi, Soonwon Choi, Renate Landig, Alexander~M Douglas, Junichi Isoya, Fedor Jelezko, Shinobu Onoda, Hitoshi Sumiya, Paola Cappellaro, et~al.
\newblock ``Quantum metrology with strongly interacting spin systems''.
\newblock \href{https://dx.doi.org/10.1103/PhysRevX.10.031003}{Physical review X {\bf 10}, 031003}~(2020).

\bibitem{rosenband2008frequency}
Till Rosenband, DB~Hume, PO~Schmidt, Chin-Wen Chou, Anders Brusch, Luca Lorini, WH~Oskay, Robert~E Drullinger, Tara~M Fortier, Jason~E Stalnaker, et~al.
\newblock ``Frequency ratio of al+ and hg+ single-ion optical clocks; metrology at the 17th decimal place''.
\newblock \href{https://dx.doi.org/10.1126/science.1154622}{Science {\bf 319}, 1808--1812}~(2008).

\bibitem{appel2009mesoscopic}
J{\"u}rgen Appel, Patrick~Joachim Windpassinger, Daniel Oblak, U~Busk Hoff, Niels Kj{\ae}rgaard, and Eugene~Simon Polzik.
\newblock ``Mesoscopic atomic entanglement for precision measurements beyond the standard quantum limit''.
\newblock \href{https://dx.doi.org/10.1073/pnas.0901550106}{Proceedings of the National Academy of Sciences {\bf 106}, 10960--10965}~(2009).

\bibitem{ludlow2015optical}
Andrew~D Ludlow, Martin~M Boyd, Jun Ye, Ekkehard Peik, and Piet~O Schmidt.
\newblock ``Optical atomic clocks''.
\newblock \href{https://dx.doi.org/10.1103/RevModPhys.87.637}{Reviews of Modern Physics {\bf 87}, 637}~(2015).

\bibitem{kaubruegger2021quantum}
Raphael Kaubruegger, Denis~V Vasilyev, Marius Schulte, Klemens Hammerer, and Peter Zoller.
\newblock ``Quantum variational optimization of ramsey interferometry and atomic clocks''.
\newblock \href{https://dx.doi.org/10.1103/PhysRevX.11.041045}{Physical review X {\bf 11}, 041045}~(2021).

\bibitem{marciniak2022optimal}
Christian~D Marciniak, Thomas Feldker, Ivan Pogorelov, Raphael Kaubruegger, Denis~V Vasilyev, Rick van Bijnen, Philipp Schindler, Peter Zoller, Rainer Blatt, and Thomas Monz.
\newblock ``Optimal metrology with programmable quantum sensors''.
\newblock \href{https://dx.doi.org/10.1038/s41586-022-04435-4}{Nature {\bf 603}, 604--609}~(2022).

\bibitem{giovannetti2006quantum}
Vittorio Giovannetti, Seth Lloyd, and Lorenzo Maccone.
\newblock ``Quantum metrology''.
\newblock \href{https://dx.doi.org/10.1103/PhysRevLett.96.010401}{Phys. Rev. Lett. {\bf 96}, 010401}~(2006).

\bibitem{metro_noise12}
Rafa{\l} Demkowicz-Dobrza{\'n}ski, Jan Ko{\l}ody{\'n}ski, and M{\u{a}}d{\u{a}}lin Gu{\c{t}}{\u{a}}.
\newblock ``The elusive heisenberg limit in quantum-enhanced metrology''.
\newblock \href{https://dx.doi.org/10.1038/ncomms2067}{Nature communications {\bf 3}, 1063}~(2012).

\bibitem{metro_noise11}
BM~Escher, Ruynet~Lima de~Matos~Filho, and Luiz Davidovich.
\newblock ``General framework for estimating the ultimate precision limit in noisy quantum-enhanced metrology''.
\newblock \href{https://dx.doi.org/10.1038/nphys1958}{Nature Physics {\bf 7}, 406--411}~(2011).

\bibitem{demkowicz2014using}
Rafal Demkowicz-Dobrza{\'n}ski and Lorenzo Maccone.
\newblock ``Using entanglement against noise in quantum metrology''.
\newblock \href{https://dx.doi.org/10.1103/PhysRevLett.113.250801}{Physical review letters {\bf 113}, 250801}~(2014).

\bibitem{sekatski2017quantum}
Pavel Sekatski, Michalis Skotiniotis, Janek Ko{\l{}}ody{\'{n}}ski, and Wolfgang D{\"{u}}r.
\newblock ``Quantum metrology with full and fast quantum control''.
\newblock \href{https://dx.doi.org/10.22331/q-2017-09-06-27}{{Quantum} {\bf 1}, 27}~(2017).

\bibitem{metro_QEC17}
Rafa\l{} Demkowicz-Dobrza\ifmmode~\acute{n}\else \'{n}\fi{}ski, Jan Czajkowski, and Pavel Sekatski.
\newblock ``Adaptive quantum metrology under general markovian noise''.
\newblock \href{https://dx.doi.org/10.1103/PhysRevX.7.041009}{Phys. Rev. X {\bf 7}, 041009}~(2017).

\bibitem{zhou2018achieving}
Sisi Zhou, Mengzhen Zhang, John Preskill, and Liang Jiang.
\newblock ``Achieving the heisenberg limit in quantum metrology using quantum error correction''.
\newblock \href{https://dx.doi.org/10.1038/s41467-017-02510-3}{Nature communications {\bf 9}, 78}~(2018).

\bibitem{huelga1997improvement}
S.~F. Huelga, C.~Macchiavello, T.~Pellizzari, A.~K. Ekert, M.~B. Plenio, and J.~I. Cirac.
\newblock ``Improvement of frequency standards with quantum entanglement''.
\newblock \href{https://dx.doi.org/10.1103/PhysRevLett.79.3865}{Phys. Rev. Lett. {\bf 79}, 3865--3868}~(1997).

\bibitem{channel_estimate21}
Sisi Zhou and Liang Jiang.
\newblock ``Asymptotic theory of quantum channel estimation''.
\newblock \href{https://dx.doi.org/10.1103/PRXQuantum.2.010343}{PRX Quantum {\bf 2}, 010343}~(2021).

\bibitem{MPS_metro13}
Marcin Jarzyna and Rafa\l{} Demkowicz-Dobrza\ifmmode~\acute{n}\else \'{n}\fi{}ski.
\newblock ``Matrix product states for quantum metrology''.
\newblock \href{https://dx.doi.org/10.1103/PhysRevLett.110.240405}{Phys. Rev. Lett. {\bf 110}, 240405}~(2013).

\bibitem{MPS_metro20}
Krzysztof Chabuda, Jacek Dziarmaga, Tobias~J Osborne, and Rafa{\l} Demkowicz-Dobrza{\'n}ski.
\newblock ``Tensor-network approach for quantum metrology in many-body quantum systems''.
\newblock \href{https://dx.doi.org/10.1038/s41467-019-13735-9}{Nature communications {\bf 11}, 250}~(2020).

\bibitem{hayashi2022global}
Masahito Hayashi, Zi-Wen Liu, and Haidong Yuan.
\newblock ``Global heisenberg scaling in noisy and practical phase estimation''.
\newblock \href{https://dx.doi.org/10.1088/2058-9565/ac5d7e}{Quantum Science and Technology {\bf 7}, 025030}~(2022).

\bibitem{QFI76}
Carl~W Helstrom.
\newblock ``Quantum detection and estimation theory''.
\newblock \href{https://dx.doi.org/10.1007/BF01007479}{Journal of statistical physics {\bf 1}, 231--252}~(1969).

\bibitem{holevo2011probabilistic}
Alexander~S Holevo.
\newblock ``Probabilistic and statistical aspects of quantum theory''.
\newblock Volume~1.
\newblock Springer Science \& Business Media. ~(2011).

\bibitem{QFI94}
Samuel~L. Braunstein and Carlton~M. Caves.
\newblock ``Statistical distance and the geometry of quantum states''.
\newblock \href{https://dx.doi.org/10.1103/PhysRevLett.72.3439}{Phys. Rev. Lett. {\bf 72}, 3439--3443}~(1994).

\bibitem{barndorff2000fisher}
Ole~E Barndorff-Nielsen and Richard~D Gill.
\newblock ``Fisher information in quantum statistics''.
\newblock \href{https://dx.doi.org/10.1088/0305-4470/33/24/306}{Journal of Physics A: Mathematical and General {\bf 33}, 4481}~(2000).

\bibitem{gill2000state}
Richard~D Gill and Serge Massar.
\newblock ``State estimation for large ensembles''.
\newblock \href{https://dx.doi.org/10.1103/PhysRevA.61.042312}{Physical Review A {\bf 61}, 042312}~(2000).

\bibitem{estimation_tech09}
Matteo G.~A. Paris.
\newblock ``Quantum estimation for quantum technology''.
\newblock \href{https://dx.doi.org/10.1142/S0219749909004839}{International Journal of Quantum Information {\bf 07}, 125--137}~(2009).

\bibitem{Shor_code95}
Peter~W. Shor.
\newblock ``Scheme for reducing decoherence in quantum computer memory''.
\newblock \href{https://dx.doi.org/10.1103/PhysRevA.52.R2493}{Phys. Rev. A {\bf 52}, R2493--R2496}~(1995).

\bibitem{knill_paritycode00}
E.~Knill, R.~Laflamme, and G.~Milburn.
\newblock ``Efficient linear optics quantum computation''~(2000).
\newblock  \href{http://arxiv.org/abs/quant-ph/0006088}{quant-ph:quant-ph/0006088}.

\bibitem{paritycode05}
T.~C. Ralph, A.~J.~F. Hayes, and Alexei Gilchrist.
\newblock ``Loss-tolerant optical qubits''.
\newblock \href{https://dx.doi.org/10.1103/PhysRevLett.95.100501}{Phys. Rev. Lett. {\bf 95}, 100501}~(2005).

\bibitem{zang2025enhancing}
Allen Zang, Tian-Xing Zheng, Peter~C. Maurer, Frederic~T. Chong, Martin Suchara, and Tian Zhong.
\newblock ``{Enhancing Noisy Quantum Sensing by GHZ State Partitioning}''~(2025).
\newblock  \href{http://arxiv.org/abs/2507.02829}{arXiv:2507.02829}.

\bibitem{squeez_93}
Masahiro Kitagawa and Masahito Ueda.
\newblock ``Squeezed spin states''.
\newblock \href{https://dx.doi.org/10.1103/PhysRevA.47.5138}{Phys. Rev. A {\bf 47}, 5138--5143}~(1993).

\bibitem{squeez_rev11}
Jian Ma, Xiaoguang Wang, C.P. Sun, and Franco Nori.
\newblock ``Quantum spin squeezing''.
\newblock \href{https://dx.doi.org/10.1016/j.physrep.2011.08.003}{Physics Reports {\bf 509}, 89--165}~(2011).

\bibitem{ulam2001spin}
Duger Ulam-Orgikh and Masahiro Kitagawa.
\newblock ``Spin squeezing and decoherence limit in ramsey spectroscopy''.
\newblock \href{https://dx.doi.org/10.1103/PhysRevA.64.052106}{Physical Review A {\bf 64}, 052106}~(2001).

\bibitem{brask2015improved}
Jonatan~Bohr Brask, Rafael Chaves, and Janek Ko{\l}ody{\'n}ski.
\newblock ``Improved quantum magnetometry beyond the standard quantum limit''.
\newblock \href{https://dx.doi.org/10.1103/PhysRevX.5.031010}{Physical Review X {\bf 5}, 031010}~(2015).

\bibitem{sorensen2001entanglement}
Anders~S S{\o}rensen and Klaus M{\o}lmer.
\newblock ``Entanglement and extreme spin squeezing''.
\newblock \href{https://dx.doi.org/10.1103/PhysRevLett.86.4431}{Physical review letters {\bf 86}, 4431}~(2001).

\bibitem{metro_rev14}
Géza Tóth and Iagoba Apellaniz.
\newblock ``Quantum metrology from a quantum information science perspective''.
\newblock \href{https://dx.doi.org/10.1088/1751-8113/47/42/424006}{Journal of Physics A: Mathematical and Theoretical {\bf 47}, 424006}~(2014).

\bibitem{pezze2009entanglement}
Luca Pezz{\'e} and Augusto Smerzi.
\newblock ``Entanglement, nonlinear dynamics, and the heisenberg limit''.
\newblock \href{https://dx.doi.org/10.1103/PhysRevLett.102.100401}{Physical review letters {\bf 102}, 100401}~(2009).

\bibitem{Lieb1972}
Elliott~H. Lieb and Derek~W. Robinson.
\newblock ``The finite group velocity of quantum spin systems''.
\newblock \href{https://dx.doi.org/10.1007/BF01645779}{Commun. Math. Phys. {\bf 28}, 251--257}~(1972).

\bibitem{review_us23}
Chi-Fang~(Anthony) Chen, Andrew Lucas, and Chao Yin.
\newblock ``Speed limits and locality in many-body quantum dynamics''.
\newblock \href{https://dx.doi.org/10.1088/1361-6633/acfaae}{Reports on Progress in Physics {\bf 86}, 116001}~(2023).

\bibitem{LRB_cor06}
S.~Bravyi, M.~B. Hastings, and F.~Verstraete.
\newblock ``Lieb-robinson bounds and the generation of correlations and topological quantum order''.
\newblock \href{https://dx.doi.org/10.1103/PhysRevLett.97.050401}{Phys. Rev. Lett. {\bf 97}, 050401}~(2006).

\bibitem{LRB_metro23}
Yaoming Chu, Xiangbei Li, and Jianming Cai.
\newblock ``Strong quantum metrological limit from many-body physics''.
\newblock \href{https://dx.doi.org/10.1103/PhysRevLett.130.170801}{Phys. Rev. Lett. {\bf 130}, 170801}~(2023).

\bibitem{bitwise_sense09}
B~L Higgins, D~W Berry, S~D Bartlett, M~W Mitchell, H~M Wiseman, and G~J Pryde.
\newblock ``Demonstrating heisenberg-limited unambiguous phase estimation without adaptive measurements''.
\newblock \href{https://dx.doi.org/10.1088/1367-2630/11/7/073023}{New Journal of Physics {\bf 11}, 073023}~(2009).

\bibitem{bitwise_sense15}
Shelby Kimmel, Guang~Hao Low, and Theodore~J. Yoder.
\newblock ``Robust calibration of a universal single-qubit gate set via robust phase estimation''.
\newblock \href{https://dx.doi.org/10.1103/PhysRevA.92.062315}{Phys. Rev. A {\bf 92}, 062315}~(2015).

\bibitem{bitwise_sense20}
Federico Belliardo and Vittorio Giovannetti.
\newblock ``Achieving heisenberg scaling with maximally entangled states: An analytic upper bound for the attainable root-mean-square error''.
\newblock \href{https://dx.doi.org/10.1103/PhysRevA.102.042613}{Phys. Rev. A {\bf 102}, 042613}~(2020).

\bibitem{echo_metro16}
Tommaso Macr\`{\i}, Augusto Smerzi, and Luca Pezz\`e.
\newblock ``Loschmidt echo for quantum metrology''.
\newblock \href{https://dx.doi.org/10.1103/PhysRevA.94.010102}{Phys. Rev. A {\bf 94}, 010102}~(2016).

\bibitem{LRB_open10}
David Poulin.
\newblock ``Lieb-robinson bound and locality for general markovian quantum dynamics''.
\newblock \href{https://dx.doi.org/10.1103/PhysRevLett.104.190401}{Phys. Rev. Lett. {\bf 104}, 190401}~(2010).

\bibitem{domino05}
Jae-Seung Lee and A.~K. Khitrin.
\newblock ``Stimulated wave of polarization in a one-dimensional ising chain''.
\newblock \href{https://dx.doi.org/10.1103/PhysRevA.71.062338}{Phys. Rev. A {\bf 71}, 062338}~(2005).

\bibitem{metro_domino21}
Atsuki Yoshinaga, Mamiko Tatsuta, and Yuichiro Matsuzaki.
\newblock ``Entanglement-enhanced sensing using a chain of qubits with always-on nearest-neighbor interactions''.
\newblock \href{https://dx.doi.org/10.1103/PhysRevA.103.062602}{Phys. Rev. A {\bf 103}, 062602}~(2021).

\bibitem{kobus2025asymptotically}
Arkadiusz Kobus and Rafał Demkowicz-Dobrzański.
\newblock ``Asymptotically optimal joint phase and dephasing strength estimation using spin-squeezed states''~(2025).

\bibitem{squeez_Norm23}
Maxwell Block, Bingtian Ye, Brenden Roberts, Sabrina Chern, Weijie Wu, Zilin Wang, Lode Pollet, Emily~J Davis, Bertrand~I Halperin, and Norman~Y Yao.
\newblock ``Scalable spin squeezing from finite-temperature easy-plane magnetism''.
\newblock \href{https://dx.doi.org/10.1038/s41567-024-02562-5}{Nature PhysicsPages 1--7}~(2024).

\bibitem{squeeze_tower22}
Tommaso Comparin, Fabio Mezzacapo, and Tommaso Roscilde.
\newblock ``Robust spin squeezing from the tower of states of u(1)-symmetric spin hamiltonians''.
\newblock \href{https://dx.doi.org/10.1103/PhysRevA.105.022625}{Phys. Rev. A {\bf 105}, 022625}~(2022).

\bibitem{squeeze_2dlocal}
Tommaso Roscilde, Filippo Caleca, Adriano Angelone, and Fabio Mezzacapo.
\newblock ``Scalable spin squeezing from critical slowing down in short-range interacting systems''.
\newblock \href{https://dx.doi.org/10.1103/PhysRevLett.133.210401}{Phys. Rev. Lett. {\bf 133}, 210401}~(2024).

\bibitem{QFI_contin16}
R.~Augusiak, J.~Ko\l{}ody\ifmmode~\acute{n}\else \'{n}\fi{}ski, A.~Streltsov, M.~N. Bera, A.~Ac\'{\i}n, and M.~Lewenstein.
\newblock ``Asymptotic role of entanglement in quantum metrology''.
\newblock \href{https://dx.doi.org/10.1103/PhysRevA.94.012339}{Phys. Rev. A {\bf 94}, 012339}~(2016).

\bibitem{QFI_contin19}
A.~T. Rezakhani, M.~Hassani, and S.~Alipour.
\newblock ``Continuity of the quantum fisher information''.
\newblock \href{https://dx.doi.org/10.1103/PhysRevA.100.032317}{Phys. Rev. A {\bf 100}, 032317}~(2019).

\bibitem{upper_triangle}
Rajendra Bhatia.
\newblock ``Pinching, trimming, truncating, and averaging of matrices''.
\newblock \href{https://dx.doi.org/10.1080/00029890.2000.12005245}{The American Mathematical Monthly {\bf 107}, 602--608}~(2000).

\bibitem{Wigner63}
E.~P. Wigner and Mutsuo~M. Yanase.
\newblock ``Information contents of distributions''.
\newblock \href{https://dx.doi.org/10.1073/pnas.49.6.910}{Proceedings of the National Academy of Sciences {\bf 49}, 910--918}~(1963).

\bibitem{skew<F}
Shunlong Luo.
\newblock ``Wigner-yanase skew information vs. quantum fisher information''.
\newblock \href{https://dx.doi.org/10.1090/S0002-9939-03-07175-2}{Proceedings of the American Mathematical Society {\bf 132}, 885--890}~(2004).

\bibitem{domino_bessel}
Benoit Roubert, Petr Braun, and Daniel Braun.
\newblock ``Large effects of boundaries on spin amplification in spin chains''.
\newblock \href{https://dx.doi.org/10.1103/PhysRevA.82.022302}{Phys. Rev. A {\bf 82}, 022302}~(2010).

\bibitem{wan2022bounds}
Kianna Wan and Robert Lasenby.
\newblock ``Bounds on adaptive quantum metrology under markovian noise''.
\newblock \href{https://dx.doi.org/10.1103/PhysRevResearch.4.033092}{Physical Review Research {\bf 4}, 033092}~(2022).

\end{thebibliography}

\end{document}